\documentclass[journal,10pt]{IEEEtran}
 
\IEEEoverridecommandlockouts
\makeatletter

\def\endthebibliography{%
\def\@noitemerr{\@latex@warning{Empty `thebibliography' environment}}%
\endlist
}
\makeatother
\usepackage{cite}
\usepackage{amsmath,amssymb,amsfonts}
\usepackage{setspace}
\usepackage{algorithm}
\usepackage{algpseudocode}

\usepackage{graphicx}
\usepackage{textcomp}
\usepackage{makecell}
\usepackage[table,xcdraw]{xcolor}
\def\BibTeX{{\rm B\kern-.05em{\sc i\kern-.025em b}\kern-.08em
	T\kern-.1667em\lower.7ex\hbox{E}\kern-.125emX}}
	
\usepackage[left=0.62in,right=0.62in,top=0.7in,bottom=0.9in]{geometry}
\makeatletter
\renewcommand{\subsection}{\@startsection{subsection}{2}{\z@}{0.8ex}{0.3ex}{\normalfont\normalsize\itshape}}
\renewcommand{\subsubsection}{\@startsection{subsubsection}{3}{\z@}{0.5ex}{0.2ex}{\normalfont\normalsize\itshape}}
\makeatother

\usepackage{setspace}
\usepackage{multirow}
\usepackage{comment}
\usepackage{tabularx}
\usepackage{booktabs}
\usepackage{graphicx}
\usepackage{subcaption}

\newcolumntype{L}{>{\centering\arraybackslash}m{4.5cm}}
\newcolumntype{K}{>{\centering\arraybackslash}m{2cm}}
\newcolumntype{R}{>{\centering\arraybackslash}m{4.5cm}}

\usepackage{xspace} 
\usepackage[acronym,nogroupskip,nonumberlist,nopostdot]{glossaries}

\newcommand{\vectr}[1]{\boldsymbol{\mathrm{#1}}}
\newcommand{\matr}[1]{\boldsymbol{\mathrm{#1}}}
	
\newcommand\cgaus[2]{\mathcal{CN}\left(#1,#2\right)}
\newcommand\rgaus[2]{\mathcal{N}\left(#1,#2\right)}
\newcommand\norm[1]{\left\lVert#1\right\rVert}
\newcommand\abs[1]{\left\lvert#1\right\rvert}
\newcommand{\eye}{\matr{I}}
\newcommand{\pilot}{\matr{\Phi}}
\newcommand{\RomanNumeralCaps}[1]{\MakeUppercase{\romannumeral #1}}

\newcommand{\bG}{\matr{G}}

\newcommand{\bY}{\matr{Y}}
\newcommand{\bH}{\matr{H}}
\newcommand{\mH}{\mathcal{H}}

\newcommand{\mL}{\mathcal{L}}
\newcommand{\bHdl}{\matr{H}_\text{DL}}
\newcommand{\bHbl}{\matr{H}_\text{BL}}
\newcommand{\bHblk}{\matr{H}_\text{BL}^k}
\newcommand{\bHblkp}{\matr{H}_\text{BL}^{k'}}
\newcommand{\bHblHat}{\hat{\matr{H}}_\text{BL}}
\newcommand{\bhc}{\matr{h}_\text{C}}
\newcommand{\bhck}{\matr{h}_\text{C}^k}

\newcommand{\bhr}{\matr{h}_\text{R}}
\newcommand{\bhrk}{\matr{h}_\text{R}^k}

\newcommand{\bW}{\matr{W}}

\newcommand{\Pmax}{P_\text{max}}
\newcommand{\bA}{\matr{A}}
\newcommand{\ba}{\vectr{a}}
\newcommand{\bB}{\matr{B}}
\newcommand{\bC}{\matr{C}}

\newcommand{\bEdl}{\matr{E}_\text{DL}}
\newcommand{\bw}{\vectr{w}}
\newcommand{\bq}{\vectr{q}}
\newcommand{\bb}{\vectr{b}}
\newcommand{\bh}{\vectr{h}}
\newcommand{\bu}{\vectr{u}}
\newcommand{\bv}{\vectr{v}}

\newcommand{\bD}{\matr{D}}
\newcommand{\bn}{\vectr{n}}
\newcommand{\bX}{\matr{X}}
\newcommand{\bx}{\vectr{x}}
\newcommand{\bz}{\vectr{z}}

\newcommand{\bZ}{\matr{Z}}

\newcommand{\mS}{\mathcal{S}}
\newcommand{\mO}{\mathcal{O}}
\newcommand{\bxbf}{\bx_\text{BF}}

\newcommand{\by}{\vectr{y}}

\newcommand{\taupref}{\tau_{p, \text{ref}}}

\newcommand{\nr}{N_\text{R}}
\newcommand{\nc}{N_\text{C}}
\newcommand{\apref}{\text{AP}_\text{ref}}
\newcommand{\refer}{\text{ref}}
\newcommand{\mP}{\mathcal{P}}
\newcommand{\Pbf}{\mathcal{P}_{\text{BF}}}
\newcommand{\Pbfp}{\mathcal{P}_{\text{BF}}^\prime}
\newcommand{\Pdli}{\mathcal{P}_{\text{DLI}}}

\newcommand{\Palpha}{\mathcal{P}_{\alpha 0}}
\newcommand{\Palphaf}{\mathcal{P}_{{\alpha 0},1}}

\newcommand{\mSce}{\mathcal{S}_\text{CE}}
\newcommand{\mSr}{\mathcal{S}_\text{R}}
\newcommand{\mSaux}{\mathcal{S}_\text{aux}}

\newcommand{\herm}{\mathsf{H}}

\newcommand{\ap}{\text{AP}}
\newcommand{\trp}{\mathsf{T}}

\newcommand{\realset}[2]{ \mathbb{R}^{#1 \times #2}  }
\newcommand{\complexset}[2]{ \mathbb{C}^{#1 \times #2}  }
\newcommand{\realR}[1]{ \mathbb{R}^{#1}  }

\newcommand{\re}[1]{\operatorname{Re}\left\{#1\right\} }

\newcommand{\floor}[1]{\left\lfloor#1\right\rfloor}
\newcommand{\im}[1]{ \operatorname{Im}\left\{#1\right\} }
\newcommand{\tr}[1]{ \operatorname{Tr}\left\{#1\right\} }
\newcommand{\ex}[1]{ \operatorname{E}\left\{#1\right\} }
\newcommand{\var}[1]{ \operatorname{Var}\left\{#1\right\} }

\newcommand{\rank}[1]{\operatorname{rank} \left( #1 \right) }

\newcommand{\diag}[1]{\operatorname{diag} \left( #1 \right) }
\newcommand{\detr}[1]{\operatorname{det} \left( #1 \right) }

\DeclareMathOperator*{\SQNR}{SQNR}

\newcommand{\aps}{\glspl{ap}\xspace}
\newcommand{\ces}{\glspl{ce}\xspace}

\newacronym{2d}{2D}{two-dimensional}
\newacronym{3d}{3D}{three-dimensional}
\newacronym{4g}{4G}{fourth generation}
\newacronym{5g}{5G}{fifth generation}
\newacronym{5gnr}{5G NR}{5G New Radio}
\newacronym{3gpp}{3GPP}{third-generation partnership project}
\newacronym{adc}{ADC}{analog-to-digital converter}
\newacronym{am}{AM}{amplitude modulation}
\newacronym{ambc}{AmBC}{ambient BC}
\newacronym{ap}{AP}{access point}
\newacronym{ar}{AR}{augmented reality}
\newacronym{aoa}{AOA}{angle-of-arrival}
\newacronym{agc}{AGC}{automatic gain control}
\newacronym{awgn}{AWGN}{additive white Gaussian noise}
\newacronym{bc}{BC}{backscatter communication}
\newacronym{bde}{BD}{backscatter device}
\newacronym{bf}{BF}{beamforming}
\newacronym{ber}{BER}{bit error rate}
\newacronym{bs}{BS}{base station}
\newacronym{bibc}{BiBC}{bistatic BC}
\newacronym{ce}{CE}{carrier emitter}
\newacronym{csi}{CSI}{channel state information}
\newacronym{cvx}{CVX}{convex optimization toolbox}
\newacronym{dac}{DAC}{digital-to-analog converter}
\newacronym{dmimo_s}{D-MIMO}{distributed MIMO}
\newacronym{dl}{DL}{downlink}
\newacronym{dmimo}{D-MIMO}{distributed multiple-input multiple-output}
\newacronym{doa}{DOA}{direction-of-arrival}
\newacronym{dli}{DLI}{direct link interference}
\newacronym{dft}{DFT}{discrete Fourier transform}
\newacronym{dtft}{DTFT}{discrete-time Fourier transform}
\newacronym{dp}{DP}{dynamic programming}
\newacronym{en}{EN}{energy neutral}
\newacronym{end}{END}{energy neutral device}
\newacronym{eirp}{EIRP}{effective isotropic radiated power}
\newacronym{etsi}{ETSI}{European Telecommunications Standards Institute}
\newacronym{evd}{EVD}{eigenvalue decomposition}
\newacronym{fdd}{FDD}{frequency-division duplexing}
\newacronym{fdma}{FDMA}{frequency-division multiple access}
\newacronym{fft}{FFT}{fast Fourier transform}
\newacronym{gs}{GS}{grid search}
\newacronym{gd}{GD}{gradient descent}
\newacronym{gsm}{GSM}{Global System for Mobile Communications}  
\newacronym{gna}{GNA}{Girvan-Newman algorithm}
\newacronym{glrt}{GLRT}{generalized log-likelihood ratio test}
\newacronym{icsi}{ICSI}{imperfect CSI}
\newacronym{iot}{IoT}{Internet-of-Things}
\newacronym{iid}{i.i.d.}{independent and identically distributed}
\newacronym{isr}{ISR}{interference-to-signal ratio}
\newacronym{ieee}{IEEE}{Institute of Electrical and Electronics Engineers}
\newacronym{kkt}{KKT}{Karush–Kuhn–Tucker}
\newacronym{los}{LoS}{line-of-sight}
\newacronym{lora}{LoRa}{long range}
\newacronym{lti}{LTI}{linear time-invariant}
\newacronym{ls}{LS}{least-squares}
\newacronym{lte}{LTE}{Long-Term Evolution}
\newacronym{lan}{LAN}{local area network}
\newacronym{lsb}{LSB}{least significant bit}
\newacronym{m}{m}{meters}
\newacronym{ml}{ML}{maximum-likelihood}
\newacronym{mse}{MSE}{mean square error}
\newacronym{mimo}{MIMO}{multiple-input multiple-output}
\newacronym{mumimo}{MU-MIMO}{multi-user \gls{mimo}}
\newacronym{miso}{MISO}{multiple-input single-output}
\newacronym{mmwave}{mmWave}{millimeter wave}
\newacronym{mmse}{MMSE}{minimum mean square error}
\newacronym{map}{MAP}{maximum a posteriori probability}
\newacronym{mrc}{MRC}{maximum-ratio combining}
\newacronym{mrt}{MRT}{maximum-ratio transmission}
\newacronym{mobc}{MoBC}{monostatic BC}
\newacronym{nmse}{NMSE}{normalized mean square error}
\newacronym{nr}{NR}{New Radio}
\newacronym{np}{NP}{Neyman-Pearson}
\newacronym{nfc}{NFC}{near-field communication}
\newacronym{nlos}{NLoS}{non-line-of-sight}
\newacronym{ofdm}{OFDM}{orthogonal frequency division multiplexing}
\newacronym{ofdma}{OFDMA}{orthogonal frequency-division multiple access}
\newacronym{ota}{OtA}{over-the-air}
\newacronym{ua}{UA}{uncertainty agnostic}
\newacronym{p1}{P1}{Phase \RomanNumeralCaps{1}}
\newacronym{p2}{P2}{Phase \RomanNumeralCaps{2}}
\newacronym{pl}{PL}{path loss}
\newacronym{pana}{PanA}{Panel A}
\newacronym{panb}{PanB}{Panel B}
\newacronym{pgd}{PGD}{projected gradient descent}
\newacronym{ple}{PLE}{path loss exponent}
\newacronym{pcsi}{PCSI}{perfect \gls{csi}}
\newacronym{papr}{PAPR}{peak-to-average power ratio}
\newacronym{pg}{PG}{path gain}
\newacronym{pdf}{pdf}{probability density function}
\newacronym{phy}{PHY}{physical layer}
\newacronym{psd}{PSD}{positive semidefinite}
\newacronym{rcs}{RCS}{radar cross section}
\newacronym{Riss}{RIS}{Reconfigurable intelligent surfaces}
\newacronym{ris}{RIS}{reconfigurable intelligent surface}
\newacronym{riss}{RIS}{reconfigurable intelligent surfaces}
\newacronym{rf}{RF}{radio frequency}
\newacronym{rfid}{RFID}{radio frequency identification}
\newacronym{rms}{RMS}{root mean square}
\newacronym{rss}{RSS}{received signal strength}
\newacronym{rv}{RV}{random variable}
\newacronym{sdma}{SDMA}{space-division multiple access}
\newacronym{sdr}{SDR}{semidefinite relaxation}
\newacronym{sdp}{SDP}{semidefinite programming}
\newacronym{si}{SI}{self-interference}
\newacronym{sic}{SIC}{successive interference cancellation}
\newacronym{sumimo}{SU-MIMO}{single-user \gls{mimo}}
\newacronym{svd}{SVD}{singular value decomposition}
\newacronym{smc}{SMC}{specular multipath component}
\newacronym{snr}{SNR}{signal-to-noise ratio}
\newacronym{sinr}{SINR}{signal-to-interference-plus-noise ratio}
\newacronym{sir}{SIR}{signal-to-interference ratio}
\newacronym{siso}{SISO}{single-input single-output}
\newacronym{simo}{SIMO}{single-input multiple-output}
\newacronym{tdoa}{TDOA}{time-difference-of-arrival}
\newacronym{toa}{TOA}{time-of-arrival}
\newacronym{tdd}{TDD}{time division multiplexing}
\newacronym{tdma}{TDMA}{time-division multiple access}
\newacronym{ue}{UE}{user equipment}
\newacronym{ul}{UL}{uplink}
\newacronym{uhf}{UHF}{ultra high frequency}
\newacronym{ula}{ULA}{uniform linear array}
\newacronym{upa}{UPA}{uniform planar array}
\newacronym{ura}{URA}{uniform rectangular array}
\newacronym{uwb}{UWB}{ultrawideband}
\newacronym{zf}{ZF}{zero-forcing}
\newacronym{qam}{QAM}{quadrature amplitude modulation}
\newacronym{qos}{QoS}{Quality of Service}
\newacronym{wlan}{WLAN}{wireless local area network}
\newacronym{wpt}{WPT}{wireless power transfer}
\newacronym{wrt}{w.r.t.}{with respect to}

\begin{document}

\title{Joint Access Point Selection and Beamforming \\ Design for Bistatic Backscatter Communication
	\thanks{
	The authors were supported by the REINDEER project of the European Union’s Horizon 2020 research and innovation program under grant agreement No.	101013425, and in part by ELLIIT and the Knut and Alice Wallenberg (KAW) Foundation. Preliminary results of this article were presented in \cite{kaplan2024reduce}.	
		}
}

\author{Ahmet Kaplan, Diana P. M. Osorio, and Erik G. Larsson}

\markboth{Submitted to IEEE Transactions on Wireless Communications}{}

\maketitle

\begin{abstract}
Future Internet-of-Things networks are envisioned to use small and cheap sensor nodes with extremely low power consumption to avoid the extensive use of batteries. To provide connectivity to a massive number of these nodes, \gls{bc} is emerging as an energy- and cost-efficient technology exploiting the reflection of radio frequency signals. 
However, challenges such as round-trip path loss and \gls{dli} between the carrier emitter and the reader limit its performance. To tackle these limitations, this paper proposes a joint access point role selection and a novel beamforming technique for bistatic \gls{bc} in a distributed multiple-input multiple-output setup. The proposed approach boosts the received backscattered energy while effectively mitigating \gls{dli}, thereby reducing the error probability. We also propose a channel estimation method tailored to operate under \gls{dli} conditions and propose a mismatch detector using estimated channel coefficients. Furthermore, we derive a closed-form expression for the probability of error for the detectors and model the quantization noise caused by \gls{dli}. Finally, comprehensive simulation results show that the proposed method with 1-bit \glspl{adc} effectively mitigates \gls{dli}, reduces the quantization noise, and enhances backscattered signal energy, achieving performance comparable to the benchmark scenario with 8-bit \glspl{adc}.
\end{abstract}

\begin{IEEEkeywords}
	Access point selection, bistatic backscatter communication, beamforming, direct link interference, internet of things
\end{IEEEkeywords}
\glsresetall

\section{Introduction}
\Gls{iot} technology spans critical sectors such as healthcare, surveillance, and agriculture. 
Passive \gls{iot} devices offer a low-cost and sustainable solution by harnessing energy from ambient sources, such as \gls{rf} signals, thus eliminating the need for batteries. A key communication technology for passive \gls{iot} devices is \gls{bc}, which enables connectivity without requiring an active circuit to generate an \gls{rf} signal \cite{galappaththige2022link}. Instead, passive devices modulate incoming \gls{rf} signals by changing their phase and amplitude to transmit information.

A \gls{bc} consists of three main components, namely \gls{ce}, reader, and \gls{bde}. The \gls{ce} transmits the carrier signal, while the reader receives the backscattered signal from the \gls{bde}. Moreover, three types of operation are possible, \gls{mobc}, \gls{bibc}, and \gls{ambc}. In \gls{mobc}, the reader and carrier emitter are collocated; while in \gls{bibc}, the \gls{ce} and reader are spatially separated, thus eliminating the need for full-duplex operation \cite{mishra2019optimal, kimionis2014increased}. \gls{ambc} is similar to \gls{bibc}, but does not have dedicated \gls{ce} and the \gls{bde} reflects incoming ambient \gls{rf} signals, such as Bluetooth or Wi-Fi \cite{hoang2020ambient} . Herein, we will focus on \gls{bibc} while addressing interested readers to find further details in comprehensive surveys on \gls{bc} in \cite{gu2024breaking, jiang2023backscatter, xu2018practical,song2021advances,rezaei2023coding,galappaththige2022link}.

\subsection{Motivation}
The performance of \gls{bc} can be significantly degraded due to two major problems, the round-trip path loss and the \gls{dli}. Due to the cascade channel, \gls{bc} experiences round-trip path loss, which weakens the backscattered signal. Moreover, the weak \gls{bc} signal received at the reader is strongly interfered by the carrier signal transmitted by the \gls{ce}, which is called \gls{dli}.  As a result, maintaining reliable communication becomes increasingly challenging as distances grow.

As a consequence of these problems, the dynamic range of the received signal (power ratio of the \gls{dli} to the \gls{bde} signal) is too high, thus high resolution \glspl{adc} are required in the reader circuitry. However, \glspl{adc} are power-hungry components, especially in multiple antenna technology, where each antenna requires its own \gls{adc}.

In light of these circumstances, \gls{bibc} offers key advantages including spatial separation of the \gls{ce} and reader, unlike \gls{mobc}, and support for transmit \gls{bf} with multiple antennas, unlike \gls{ambc}.
For the aforementioned reasons, this paper investigates the benefits of distributed \gls{mimo} technology to enhance \gls{bibc} by deploying multiple \glspl{ap} across a geographic area. With spatial distributed \glspl{ap}, the round-trip path loss can be reduced due to macro diversity gain, thus allowing for extended range, increased data rates, improved interference management, and scalability for passive \gls{iot} devices.

\subsection{Related Work}
This subsection reviews related literature considering two main categories of works, i) \gls{bibc} with multi-antenna technology and ii) \gls{dli} cancellation in \gls{bibc} systems.

\subsubsection{\gls{bibc} with multi-antenna technology}
The works in \cite{galappaththige2024cell,qu2022channel,rezaei2023time} propose novel channel estimation algorithms for multi-antenna \gls{bibc}. Particularly, the work in~\cite{galappaththige2024cell}  optimizes the sum rate by employing \gls{bf} and multiple \glspl{bde}.
In \cite{qu2022channel}, the \gls{snr} in a multi-antenna reader is maximized using the channel estimates in a \gls{bibc} setup with multi-antenna \gls{ce} and a single antenna tag.
In \cite{rezaei2023time}, the proposed approach allows for the estimation of the channels of several \glspl{bde} in one shot in a \gls{bibc} setup with a multi-antenna reader.

Rate optimization is explored in~\cite{sacarelo2021bistatic, duan2017achievable, zargari2024deep} for the \gls{bibc} setup with multi-antenna \gls{ce} and multi-antenna reader. For instance,  \cite{sacarelo2021bistatic} proposes the maximization of the minimum rate of multiple \glspl{bde}.
The achievable rate with a \gls{ce} transmitting information to a reader is investigated in \cite{duan2017achievable}. 
In \cite{zargari2024deep}, deep reinforcement learning technique is employed to maximize the \glspl{bde} sum rate.

Power beacon placement to extend coverage in \gls{bibc} is studied in \cite{jia2021power}, while \cite{han2017wirelessly} examines network capacity and coverage with multiple \glspl{bde} and power beacons (\glspl{ce}). Moreover, space-time code design for \gls{bibc} and \gls{mobc} systems with multi-antenna reader, \gls{ce}, and \gls{bde} is investigated in \cite{luan2021better}.

Our previous work, in~\cite{kaplan2024access}, explores \gls{ap} role selection for a \gls{bibc} setup with distributed \gls{mimo}. Different from that work, 
this work introduces a channel estimation algorithm, proposes a \gls{bf} design, and addresses the \gls{dli} problem.

\subsubsection{DLI cancelation in \gls{bibc}}
The works in~\cite{varshney2017lorea, lopez2023designing, li2019capacity,tao2021novel,li2024code,kan2023differential,biswas2021direct,li2019adaptive, rostami2020redefining} investigate the \gls{dli} cancellation in \gls{bibc}. 
Particularly, in \cite{varshney2017lorea, lopez2023designing} and \cite{li2019capacity}, the carrier frequency of the backscattered signal is shifted at the \gls{bde} to separate the \gls{dli} and the weak \gls{bc} signal in the reader. This technique may increase the complexity of the \gls{bde} and requires additional frequency bands. 

Moreover, in \cite{tao2021novel} and \cite{li2024code}, coding in the \gls{bde} is used to cancel the \gls{dli} in the reader in a \gls{siso} \gls{bibc} setup, but the interference cancellation occurs after the \gls{adc} in the digital domain, thus requiring a high resolution \gls{adc}. In \cite{kan2023differential}, the \gls{dli} is first estimated in the digital domain and then subtracted from the received signal in a \gls{siso} \gls{bibc} setup, thus still requiring a high resolution \gls{adc}.

In \cite{biswas2021direct}, the effect of the \gls{dli} on the communication range is investigated in \gls{bibc} and \gls{ambc}.
In \cite{li2019adaptive}, the effect of imperfect \gls{sic} is investigated in a \gls{siso} \gls{bibc} setup with a hybrid \gls{bde} supporting both passive and active operations. In \cite{chen2024multi}, the \gls{dli} is first estimated and then subtracted in the analog domain in a \gls{siso} \gls{bibc}. 

In our previous work in~\cite{kaplan2023direct}, we propose a method to cancel the \gls{dli} in a \gls{bibc} setup using a multi-antenna \gls{ce} and reader. That work assumes no prior information about the \gls{bde} and focuses solely on \gls{dli} cancellation without considering the enhancement of the backscattered signal’s power. In \cite{kaplan2024reduce}, we address both \gls{dli} cancellation and power focusing to the \gls{bde} in a \gls{bibc} setup with multi-antenna \gls{ce} and reader. However, those studies involve only two \glspl{ap}, namely a \gls{ce} and a reader.

The summary of the literature on the \gls{dli} cancelation in \gls{bibc} is given in Table \ref{tab:literature_summary}.

\begin{table}[tbp]
\centering
\caption{Comparison of Related Works on \gls{dli} Cancelation}
\label{tab:literature_summary}
\resizebox{0.5\textwidth}{!}{%
\begin{tabular}{|c|>{\centering\arraybackslash}m{1.2cm}|>{\centering\arraybackslash}m{0.9cm}|>{\centering\arraybackslash}m{0.9cm}|>{\centering\arraybackslash}m{1.2cm}|>{\centering\arraybackslash}m{3.7cm}|}
\hline
\textbf{Ref} & \textbf{\begin{tabular}[c]{@{}c@{}}Dist. \\ MIMO\end{tabular}} & \textbf{\begin{tabular}[c]{@{}c@{}}Mult.\\ BD\end{tabular}} & \textbf{\begin{tabular}[c]{@{}c@{}}Chn.\\ Est.\end{tabular}} & \textbf{\begin{tabular}[c]{@{}c@{}}AP Role\\ Selection\end{tabular}} & \textbf{\begin{tabular}[c]{@{}c@{}}Novel DLI\\ Cancelation Algorithm\end{tabular}} \\ \hline
\cite{kaplan2024reduce} &  &  &  &  & Transmit beamforming \\ \hline
\cite{zargari2024deep} &  & \checkmark &  &  & Imperfect SIC considered \\ \hline
\cite{varshney2017lorea, lopez2023designing} &  & &  &  &  Frequency shift in \gls{bde} \\ \hline
\cite{li2019capacity} &  & \checkmark &  &  &  Frequency shift in \gls{bde} \\ \hline
\cite{tao2021novel, li2024code} &  &  &  &  & Coding in BD, requires high-res. ADC \\ \hline
\cite{kan2023differential} &  &  &  &  & Estimated and subtracted, requires high-res. ADC \\ \hline
\cite{biswas2021direct} &  &  &  &  & DLI effect analysis \\ \hline
\cite{li2019adaptive} &  &  &  &  & Imperfect SIC considered \\ \hline
\cite{chen2024multi} &  &  &  &  & Canceled in analog domain \\ \hline
\cite{kaplan2023direct} &  &  & \checkmark &  & Transmit beamforming \\ \hline
\textbf{This work} & \checkmark & \checkmark & \checkmark & \checkmark & Transmit beamforming and AP role selection \\ \hline
\end{tabular}%
}
\end{table}

\subsection{Contributions}
Different from the aforementioned works, this paper addresses the \gls{dli} cancellation while increasing the received backscattered energy by jointly performing \gls{ap} selection and \gls{bf} design in a distributed \gls{mimo} setup. To the best of our knowledge, no existing literature addresses the same setup and problem. Our contributions can be summarized as follows:  

\begin{itemize}
	\item For different scenarios, optimization problems are defined to effectively address the  \gls{dli} and round-trip path loss in \gls{bibc}. We propose the joint \gls{ap} selection and \gls{bf} design to mitigate the \gls{dli}, enhance the received backscattered energy, and minimize the probability of error.
	\item A channel estimation algorithm is designed to operate effectively under \gls{dli} conditions in a \gls{bibc} setup with distributed \gls{mimo}.
	\item We propose detectors for both perfect and imperfect \glspl{csi} and derive closed-form expressions for the probability of error.
	\item The quantization noise due to \gls{dli} is modeled, and the performance of the joint \gls{ap} selection and \gls{bf} algorithm is analyzed under quantization noise.
    \item 
    Simulation results demonstrate the superior performance of the proposed algorithms in canceling \gls{dli}, enhancing backscattered energy, and improving error probability, even with $1$-bit \glspl{adc}, compared to benchmark methods that use higher-resolution \glspl{adc} and design beamforming vectors without addressing \gls{dli}.
\end{itemize}

Note that, in \cite{kaplan2024reduce}, perfect channel knowledge is assumed, the \gls{dli} constraint is not defined per receive antenna, and neither \gls{ap} selection nor quantization noise modeling is addressed. In contrast, this paper introduces channel estimation, quantization noise modeling, a per-antenna \gls{dli} constraint, and a joint \gls{ap} selection and \gls{bf} algorithm for systems with multiple \glspl{ap}.

The remaining part of this paper is organized as follows. Sections \ref{sec:system_model}, \ref{sec:chn_est}, and \ref{sec:detector} present the system model, the proposed channel estimation algorithm, and the detector design, respectively. Sections \ref{sec:problems}, \ref{sec:all_bf_sol}, and \ref{sec:ap_selection} define the optimization problems for maximizing the received backscattered energy under different scenarios, describe the proposed \gls{bf} algorithm, and introduce the \gls{ap} selection strategy, respectively. Complexity analysis and numerical results are provided in Sections \ref{sec:complexity} and \ref{sec:numerical_results}, respectively. Finally, Section \ref{sec:conclusion} concludes the paper.

\textbf{Notation:} 
$(\cdot)^\trp$, $(\cdot)^\herm$, $(\cdot)^*$, $\re{\cdot}$, and $\im{\cdot}$ denote transpose, Hermitian transpose, conjugate, and real and imaginary parts, respectively. For a set $\mS$, $\abs{\cdot}$ denotes cardinality, and for a scalar, absolute value.
$\tr{\cdot}$ is the trace operator. Bold letters represent matrices (capital) and vectors (lowercase), while italic letters represent scalars. $\norm{\bX}$ and $\norm{\bx}$ are the Frobenius and Euclidean norms, respectively. $[\bx]_{i:j}$ represents elements from the $i$-th to the $j$-th position of the vector $\bx$, while $[\bX]_{i,j}$ is the element in the $i$-th row and $j$-th column of the matrix $\bX$. The $i$-th element of $\bx$ is $x_i$. Note that all vectors are column vectors. 
Expected value and variance of a vector are $\ex{\bx}$ and $\var{\bx}$. 
A random variable \(x \sim \mathcal{U}(a, b)\) follows a uniform distribution on the interval \([a, b]\).
The operator $\otimes$ denotes Kronecker product, and $\eye_{M}$ is $M \times M$ identity matrix.
The floor function is $\floor{\cdot}$, and $\forall$ and $\exists$ denote for all and there exists, respectively.
$\mathbb{C}$ and $\mathbb{R}$ are the complex and real fields.

\section{System Model}\label{sec:system_model}

\begin{figure}[tbp]
	\centering
	\includegraphics[width = 0.8\linewidth]{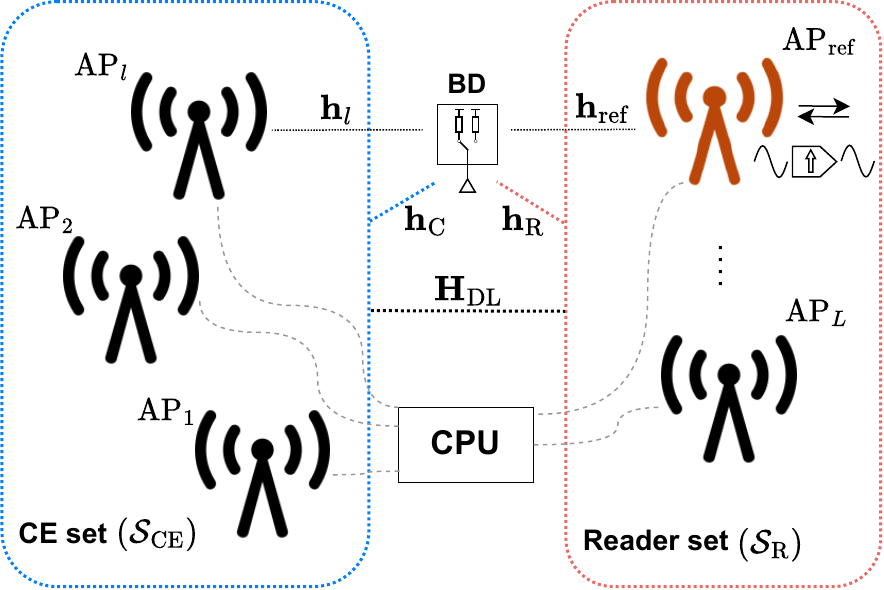}
	\caption{System model of multi-antenna \gls{bibc}.}
	\label{fig:System_Model}
\end{figure}

Fig. \ref{fig:System_Model} illustrates the system model of \gls{bibc} in a distributed \gls{mimo} implementation. In that system, there are $L$ \aps with multiple antennas and a \gls{bde} with a single antenna. The \aps are connected to a central processing unit via a fronthaul network and can operate as a reader or a \gls{ce}. Although the distributed setup requires synchronization, calibration, and fronthaul connectivity, these features are essential for emerging cellular networks, which our setup relies on. The \glspl{ap} involved in \gls{bc} are assumed to be part of this coordinated infrastructure, and \gls{bc} operations reuse existing coordination mechanisms.

The set of \aps is denoted by $\mathcal{L} = \{1,2,\dotsc,L\}$, where $\ap_l$ has $M_l$ antennas with $l \in \mathcal{L}$. 
The set  $\mS=\{\mS_\text{CE}, \mS_\text{R}\}$ denotes the partition of \aps, where $\mS_\text{CE}$ and $\mS_\text{R}$ are the sets of \ces and readers, respectively, and the sets include the indices of the \aps. The total number of antennas in the \gls{ce} set and the reader set are $\nc = \sum_{l \in \mSce} M_l$ and $\nr = \sum_{l \in \mSr} M_l$, respectively.

We assume that one of the \glspl{ap}, denoted as the reference \gls{ap} ($\apref$), is equipped with $M_\refer$ antennas, high-resolution \glspl{adc}, and operates in full-duplex mode. $\apref$ is used to enhance the \gls{bc} performance and simplify the channel estimation process.
The channels are assumed frequency-flat and defined as follows:
\begin{itemize}
    \item $\bh_l \in \complexset{M_l}{1}$: channel between $\ap_l$ and the \gls{bde},
    \item $\bh_\refer \in \complexset{M_\refer}{1}$: channel between $\apref$ and the \gls{bde}
    \item $\bHdl \in \complexset{\nr}{\nc}$: channel between \ces and readers,
	\item $\bhc \in \complexset{\nc}{1}$: channel between \ces and the \gls{bde},
	\item $\bhr \in \complexset{\nr}{1}$: channel between readers and the \gls{bde},
	\item $\bHbl = \bhr \bhc^\trp$: cascade channel of \gls{bc},
\end{itemize}
where $\bHdl, \bhc,$ and $\bhr$ are formed by assembling $\bh_l$s.

\gls{bde} transmits information by toggling the reflection coefficient $\gamma_j$ between two states, i.e., $1$ and $-1$, by adjusting the load connected to its antenna.
We aim to reduce the \gls{dli} and increase the received backscatter energy in the readers to detect the \gls{bde} information bits, i.e., the reflection coefficient sequence. 

Note that this work focuses on the single \gls{bde} scenario to provide clear analytical insights. However, the proposed algorithms can be extended to support multiple \glspl{bde} if all \glspl{bde} use mutually orthogonal reflection coefficient sequences to transmit information bits. In that case, the channels related to $k$-th \gls{bde} are defined using superscript as $\bh_{(\cdot)}^k$ and $\bH_{\text{BL}}^k$, where $k \in \{1,\dotsc, K\}$ and $K$ is the number of \glspl{bde}. For example, $\bh_{\text{R}}^k$ is the channel between readers and the $k$-th \gls{bde}.

\section{Channel Estimation Algorithm} \label{sec:chn_est}
In this section, a channel estimation algorithm is proposed to estimate the channels between each \gls{ap} and \gls{bde}. We assume the direct-link channels between \glspl{ap} are known because their estimation is straightforward \cite{kaplan2023direct}. $\apref$ has a sufficiently high dynamic range; therefore, the quantization noise in $\apref$ is negligible and is not considered in this section.

Initially, all \glspl{ap} transmit a pilot signal $J^\prime$ times in orthogonal time slots, and the backscattered pilot signals are received by $\apref$. Then, when $\ap_l$ sends the orthogonal pilot signal $J^\prime$ times, the $j^\prime-$th received signal in $\apref$ can be written as
\begin{equation} \label{eq:y_in_chn_est}
	\bY_{l,j^\prime} = \gamma_{j^\prime} \bh_\refer \bh_l^\trp  \pilot_l + \bW_{l,j^\prime},
\end{equation}
where $\bY_{l,j^\prime} \in \complexset{M_\refer}{\tau_{p,l}}$, $l \in \{1,2,\dotsc,L\}$, and $j^\prime \in \{1,\dotsc,J^\prime\}$. 
The reflection coefficient of the \gls{bde}, $\gamma_{j^\prime} \in \mathbb{C}$, is known and pre-determined during the channel estimation phase.
$\pilot_l \in \complexset{M_l}{\tau_{p,l}}\ ( \tau_{p,l} \geq M_{l})$ denotes the pilot matrix, with $\pilot_l \pilot_l^\herm = \alpha_l \eye_{M_l}$, and $\alpha_l=P_\text{max} \tau_{p,l} / M_l$, where $P_\text{max}$ is the transmitted energy in one slot.
The matrix $\bW_{l,j^\prime} \in \complexset{M_\refer}{\tau_{p,l}}$ is the additive Gaussian noise and all its elements are \gls{iid} $\cgaus{0}{1}$. 

The pilot matrix can be designed using various orthogonal structures such as \gls{dft} matrix, Hadamard matrix, or Zadoff-Chu sequences. In the case of low resolution \glspl{dac}, the Hadamard matrix is particularly suitable due to its binary-valued entries ($\pm 1$), which align well with the limited amplitude and phase resolution of the transmitter.

For simplicity, we omit the received direct-link signal from $\ap_l$ to $\apref$ in~\eqref{eq:y_in_chn_est}, since the direct-link channel is known and can be subtracted from the received signal at $\apref$ \cite{galappaththige2024cell}. In addition,  $\apref$ can operate in full-duplex mode and suppress the self-interference, when it sends the pilot signal $\pilot_\refer$ and receives $\bY_{\refer,j^\prime}$ \cite{mishra2019optimal}. 

Therefore, the channel estimation problem based on the \gls{ml} criterion can be written as follows
\begin{equation} \label{eq:ml_chn_est}
\begin{split}
\hat{\bh}_\refer, \hat{\bh}_{l}	&= \arg\min_{\bh_\refer, \bh_{l}} 
    \biggl( 
    \sum_{j' = 1}^{J'} \norm{\bY_{\refer,j^\prime} - \gamma_{j^\prime} \bh_\refer \bh_\refer^\trp  \pilot_\refer}^2 \\
    &\quad +  
    \sum_{l \in \mL \backslash  \refer}  \sum_{j' = 1}^{J'} \norm{\bY_{l,j^\prime} - \gamma_{j^\prime} \bh_\refer \bh_l^\trp  \pilot_l}^2
     \biggl),
     \end{split}
\end{equation}
where we separate the signals received when $\apref$ transmits, $\bY_{\refer,j^\prime}$, and the received signals when the remaining \glspl{ap} transmit, $\bY_{l,j^\prime}$. The objective function in \eqref{eq:ml_chn_est} is denoted by $f(\bh_\refer, \bh_{l})$.
We solve the channel estimation problem using an iterative algorithm, with the following steps.

\subsection{Step 1: Channel Estimation Between the \gls{bde} and $\text{AP}_{\mathrm{ref}}$} \label{sec:chn_est_step_1}
In this step, $\apref$ first sends the pilot signal and then estimates the cascade \gls{bc} channel using the backscattered pilot signal. After this, $\bh_\refer$ is estimated using the cascade channel estimate and the algorithm provided in \cite{mishra2019optimal}.

When $\apref$ transmits the pilot signal, the backscattered pilot signal received at $\apref$ during the $j^\prime$-th time slot is expressed~as
\begin{equation} \label{eq:rx_sgn_at_ref_in_chn_est}
	\bY_{\refer,j^\prime} =\gamma_{j^\prime} \bh_\refer \bh_\refer^\trp  \pilot_\refer+ \bW_{\refer,j^\prime}.
\end{equation}

The \gls{ls} estimate of the cascade channel $\bH_\refer=\bh_\refer \bh_\refer^\trp$ is given by
\begin{equation} \label{eq:est_of_Href}
	\hat{\bH}_\refer = \sum_{j^\prime=1}^{J^\prime} \frac{1}{J^\prime \gamma_{j^\prime}}\bY_{\refer,j^\prime} \pilot_\refer^\herm (\pilot_\refer \pilot_\refer^\herm)^{-1}.
\end{equation}
For the estimation of $\bh_\refer$, we follow the same steps as in \cite{mishra2019optimal}, where a real symmetric matrix, $\breve{\bG} \in \realset{2M_{\refer}}{2M_{\refer}}$, with real eigenvalues is formed by defining $\breve{\bH} \triangleq (\hat{\bH}_{\refer}^\herm + \hat{\bH}_{\refer}^*)/2$ and expressing it as
\begin{equation}
	\breve{\bG} \triangleq
	\begin{bmatrix}
		\re{\breve{\bH}} & -\im{\breve{\bH}} \\
		-\im{\breve{\bH}} & -\re{\breve{\bH}}
	\end{bmatrix}.
\end{equation} 
The real and imaginary parts of $\bh_{\refer}$ are then estimated as
\begin{equation}\label{eq:evd}
	\allowdisplaybreaks
	\begin{split}
		\begin{bmatrix}
			\re{\hat{\bh}_\refer} \\ \im{\hat{\bh}_\refer} 
		\end{bmatrix}
		\triangleq
		\pm
		\sqrt{\lambda_{\breve{G}_1}} \frac{\bq_{\breve{G}_1}}{\norm{\bq_{\breve{G}_1}}} \in \realset{2M_\refer}{1},
	\end{split}
\end{equation}
where $\lambda_{\breve{G}_1}$ is the maximum eigenvalue of $\breve{\bG}$, and $\bq_{\breve{G}_1}$ is the corresponding eigenvector. Finally, the estimate of $\bh_\refer$ is
\begin{equation} \label{eq:est_of_href}
	\hat{\bh}_\refer = \re{\hat{\bh}_\refer} + j \im{\hat{\bh}_\refer}.
\end{equation}
Note that there can be 180-degree phase ambiguity in the estimate of $\bh_\refer$, but this does not affect the performance as shown in the next subsection.

\subsection{Step 2: Channel Estimation Between the \gls{bde} and the Remaining \glspl{ap}} \label{sec:estimae_of_hl}
In this step, we estimate $\bh_l\ (\forall l \in \mL \backslash \refer)$ using $\hat{\bh}_\refer$ and $\bY_{l,j^\prime}$.
First, the cascade \gls{bc} channel is estimated using the \gls{ml} algorithm as follows
\begin{equation}\label{eq:chn_est_step_2}
	\widehat{\bh_\refer \bh_l^\trp} = \sum_{j^\prime=1}^{J^\prime} \frac{1}{J^\prime \gamma_{j^\prime}}\bY_{l,j^\prime} \pilot_l^\herm (\pilot_l \pilot_l^\herm)^{-1}.
\end{equation}
Next, the estimate of $\bh_l$ is calculated as 
\begin{equation} \label{eq:est_of_h_l}
\hat{\bh_l}^\trp = \frac{\hat{\bh}_\refer^\herm}{\norm{\hat{\bh}_\refer}^2} \widehat{\bh_\refer \bh_l^\trp},
\end{equation}
where $l \in \mL \backslash \refer$.
The estimate of each $\bh_l$ is exposed to the same phase change coming from $\hat{\bh}_\refer$, as seen in~\eqref{eq:est_of_h_l}. Since there is no phase mismatch between different \glspl{ap}, multiple \glspl{ap} can be used for coherent operations.

\begin{algorithm}[tpb]
	\caption{Gradient Descent for Estimating $\mathbf{h}_\refer$}
    \label{alg:gradient_descent}
	\begin{algorithmic}[1] 
		\Require $\bZ_{l,j^\prime}$, initial estimate $\hat{\bh}_\refer$, estimates $\hat{\mathbf{h}}_l$, learning rate $\alpha_{lr}$, maximum iteration $T$, and tolerance $\epsilon$.
		\State \textbf{Initialize:} Set $\mathbf{h}_\refer^{(0)} = \hat{\bh}_\refer$ and $\nabla \mathcal{P}(\bh_\refer^{(-1)})=0$.
		\For{$t \gets 1 \text{ to } T$}
		\State Compute the gradient of the objective function: $\nabla \mathcal{P}(\bh_\refer^{(t-1)})$.	
		\State 
            $
		\bh_\refer^{(t)} = \mathbf{h}_\refer^{(t-1)} - \alpha_{lr}  \left( \nabla \mathcal{P}(\bh_\refer^{(t-1)}) \right)^*
		$ \cite[Ch.~8]{sayed2011adaptive}.
		\State 
		$
		\text{If } \norm{\nabla \mathcal{P}(\bh_\refer^{(t-1)}) - \nabla \mathcal{P}(\bh_\refer^{(t-2)})}^2 < \epsilon, \text{ stop.}
		$
		\EndFor
		\Ensure Estimated channel vector $\hat{\mathbf{h}}_\refer = \mathbf{h}_\refer^{(t)}$.
	\end{algorithmic}
\end{algorithm}

 \subsection{Step 3: Refinement of the Estimation of $\bh_{\text{\textnormal{ref}}}$}
 \label{sec:chn_est_step_3}
In Step $1$, $\bh_\refer$ is estimated using $\bY_{\refer,j^\prime}$. However, the initial estimate of $\bh_\refer$ can be refined using 
all received pilot signals, i.e., $\bY_{l,j^\prime}$. Consequently, the initial estimate of $\bh_l$ in Step 2 can also be refined using the improved estimate of $\bh_\refer$.

Once all pilot signals are received, and we have the estimation of $\hat{\bh}_l$s, we can refine the estimation of $\bh_\refer$ by formulating the estimation problem as \cite{kaplan2023direct}
\begin{equation} \label{eq:chn_est_gd}
\resizebox{0.7\linewidth}{!}{$
    \begin{split}
    \hat{\bh}_\refer	&= \arg\min_{\bh_\refer} 
    \biggl( 
    \sum_{j' = 1}^{J'} \norm{\frac{\bY_{\refer,j^\prime} \pilot_\refer^\herm}{\gamma_{j^\prime} \alpha_\refer} -  \bh_\refer \bh_\refer^\trp  }^2 \\ &\quad +  
     \sum_{l \in \mL \backslash  \refer}  \sum_{j' = 1}^{J'} \norm{\frac{\bY_{l,j^\prime} \pilot_l^\herm}{\gamma_{j^\prime} \alpha_l} -  \bh_\refer \hat{\bh}_l^\trp }^2
     \biggl).
     \end{split}
     $}
\end{equation}
To simplify the notation we define 
$
    \bZ_{l,j^\prime} \triangleq \bY_{l,j^\prime} \pilot_{l}^\herm / (\gamma_{j^\prime} \alpha_{l}).
$
This estimation problem can be solved by using \gls{gd} with the initial estimate of $\bh_\refer$ from Step $1$ as an initial vector. 
The details are given in Algorithm \ref{alg:gradient_descent},
where the gradient of the objective function in~\eqref{eq:chn_est_gd} with respect to $\bh_\refer$
is calculated as \cite{brandwood1983complex}
\begin{equation}\label{eq:gradient}
    \begin{split}
    \nabla \mathcal{P}(\bh_\refer) &= \sum_{j' = 1}^{J'} 2 \norm{\bh_\refer}^2 \bh_\refer^* - (\bZ_{\refer,j^\prime}^* + \bZ_{\refer,j^\prime}^\herm) \bh_\refer \\ 
    &\quad +  \sum_{l \in \mL \backslash  \refer}  \sum_{j' = 1}^{J'} \norm{\hat{\bh}_l}^2 \bh_\refer^* - \bZ_{l,j^\prime}^* \hat{\bh_l}.
    \end{split}
\end{equation}

 \subsection{Step 4: Iteration of Steps $2$ and $3$}
 \label{sec:chn_est_step_4}
To improve the channel estimates, we repeatedly perform Steps $2$ and $3$ until either the squared norm of the difference between the last two estimates of $\bh_\refer$ falls below a small threshold or a pre-determined maximum number of iterations, $\zeta_{\text{chn}}$, is reached. To ensure robustness against divergence, if the maximum iteration limit is reached, the final estimates $\hat{\bh}_\refer$ and $\hat{\bh}_l$ are selected from the iteration $i^\star$ that gives the minimum value of the objective function $f(\hat{\bh}_\refer^{(i)}, \hat{\bh}_l^{(i)})$, where $i \in \{0, 1, \dotsc, \zeta_\text{chn}\}$.

Algorithm~\ref{alg:channel_estimation} summarizes the channel estimation algorithm. The total pilot overhead is 
$
J' \sum_{l=1}^{L} \tau_{p,l}
$
symbols and it scales linearly with the number of APs.
Note that the convergence of the Algorithm \ref{alg:channel_estimation} to a global optimal point cannot be guaranteed due to the non-convex nature of the objective function in \eqref{eq:ml_chn_est}. However, the algorithm will converge to satisfactory estimates within a few iterations in practice when the \gls{snr} is moderate to high due to the accurate initial estimates.

\begin{algorithm}[tpb]
\caption{Proposed Channel Estimation Procedure}
\label{alg:channel_estimation}
\begin{algorithmic}[1]
\Require $\bY_{l,j'}$, $\pilot_l$, $\gamma_{j'}$, $\epsilon$, $\zeta_{\text{chn}}$
\State Estimate initial $\hat{\bh}_\refer^{(0)}$ using $\bY_{\refer,j^\prime}$ as in Step 1
\State For each AP $l \in \mL \backslash \refer$ estimate $\hat{\bh}_l^{(0)}$ using $\hat{\bh}_\refer^{(0)}$ as in~\eqref{eq:est_of_h_l}
\State Set iteration counter $i \gets 0$
\Repeat
    \State $i \gets i + 1$
    \State Estimate $\hat{\bh}_\refer^{(i)}$ using $\hat{\bh}_l^{(i-1)}$ and Algorithm~\ref{alg:gradient_descent}
    \State For each AP $l \in \mL \backslash \refer$
     estimate $\hat{\bh}_l^{(i)}$ using $\hat{\bh}_\refer^{(i)}$ as in~\eqref{eq:est_of_h_l}  
\Until{$\|\hat{\bh}_\refer^{(i)} - \hat{\bh}_\refer^{(i-1)}\|^2 < \epsilon$ or $i = \zeta_{\text{chn}}$}
 \If{$i = \zeta_{\text{chn}}$} 
    \State \textbf{Output:} Estimates $\hat{\bh}_\refer$, $\hat{\bh}_l$ minimizing $f(\hat{\bh}_\refer^{(i)}, \hat{\bh}_l^{(i)})$ over $i \in \{0, \dotsc, \zeta_{\text{chn}}\}$
     \Else 
     \State \textbf{Output:} $\hat{\bh}_\refer = \hat{\bh}_\refer^{(i)}$, $\hat{\bh}_l = \hat{\bh}_l^{(i)}$
     \EndIf
\end{algorithmic}
\end{algorithm}

\subsection{Special Case, $M_{\text{\textnormal{ref}}}=1$}
For $M_\refer=1$, the estimation of $h_\refer$ can be simplified to 
\begin{equation}
\resizebox{0.7\linewidth}{!}{$
    \hat{h}_\refer = \sqrt{\hat{H}_\refer} = \sqrt{\abs{\hat{H}_\refer} e^{j\varphi}} = \sqrt{\abs{\hat{H}_\refer}} e^{j\varphi/2},
    $}
\end{equation} 
where $\hat{H}_\refer$, a complex scalar, is calculated using~\eqref{eq:est_of_Href}. Next, the estimate of $\bh_l$ is calculated as in Section \ref{sec:estimae_of_hl}. Using~\eqref{eq:gradient}, one can show that $\nabla \mathcal{P}(\hat{h}_\refer) = 0$, making these estimates optimal. Therefore, there is no need to apply Step 3 and Step 4 of the proposed channel estimation algorithm in this special case.  

\subsection{Multiple \glspl{bde} Case}
In the case of multiple \glspl{bde}, the received pilot signal in~\eqref{eq:y_in_chn_est} contains the superposition of all reflected signals from different \glspl{bde}, $\sum_k \gamma_{j^\prime,k} \bh_{\text{ref}}^k (\bh_{l}^k)^\trp  \pilot_l$. To estimate the channels associated with the $k$-th \gls{bde}, the channel estimation procedure must be repeated using the reflection coefficient sequence of the $k$-th \gls{bde}, i.e., $\gamma_{j',k}$. Note that $J' \geq K$ to satisfy the orthogonality condition. For example, to estimate $\bh_{\text{ref}}^k$ and $\bh_{\text{ref}}^k (\bh_{l}^k)^\trp$, $\gamma_{j',k}$ should be used in the computations of \eqref{eq:est_of_Href} and \eqref{eq:chn_est_step_2}, respectively. Similarly, in \eqref{eq:gradient}, $\gamma_{j',k}$ must be used to estimate $\bh_{\text{ref}}^k$. In summary, the channel estimation steps can be repeated per \gls{bde} due to the orthogonal reflection coefficients.

\section{Proposed Detector and Probability of Error} \label{sec:detector}
In this section, we first model the quantization error in the output of \glspl{adc} in the reader circuitry. Then, we define the hypothesis testing problem considering quantization noise and propose the \gls{map} detector to detect the information bits from the \gls{bde} for both imperfect and perfect \gls{csi} cases. In addition, we derive the probability of error expressions for the proposed detector under both cases.

\subsection{Modeling of Quantization Noise}
We assume that each antenna has two \glspl{adc}, one for the in-phase component and one for the quadrature-phase component.
Each \gls{adc} uses a uniform mid-rise quantization. The nonlinear quantization function $Q(y)$ maps the real-valued input $y \in \realR{}$ to one of the quantization levels, and $\Delta$ which is the the quantization step size is adjusted as 
\begin{equation}\label{eq:step_size}
\Delta = \sqrt{\ex{y^2}}/2^{b-1}.
\end{equation}
The number of quantization levels is $2^b$, where $b$ is the number of bits in \glspl{adc}.

The quantization noise is defined as 
$
	n = Q(y) - y.
$
If $y$ is Gaussian, $n$ is uniformly distributed with the variance approximated as $\ex{n^2} = \sigma_n^2 \approx \Delta^2/12$ \cite{sarajlic2017low}.
For analytic tractability, we assume that the quantization noise follows a zero-mean Gaussian distribution, denoted as $\mathcal{N}(0, \sigma_n^2)$. When the input to the quantization function is a vector, the function is applied element-wise, and the elements of the quantization noise are assumed to be independently distributed $\mathcal{N}(0, \sigma_n^2)$. 
While the quantization noise may exhibit spatial correlation, as each receive antenna observes a different linear combination of the same transmitted signal, we follow i.i.d. Gaussian assumption to simplify the analysis \cite{bjornson2018hardware,jacobsson2017throughput}. This approximation is also accurate in the low-\gls{snr} region.

For \gls{bc}, we model the quantizer input $y$ as 
$
	y = x + i + w,
$
where $x$ is the desired backscattered signal, $i$ is the \gls{dli} originating from the direct path between the transmitter and receiver, and $w$ is the additive noise term.
The signal-to-quantization noise ratio is defined as 
\begin{equation}\label{eq:sqnr}
	\SQNR = \frac{\ex{x^2}}{\ex{n^2}} \approx \frac{3 \times 2^{2b} \ex{x^2}}{\ex{y^2}},
\end{equation}
which increases by $6.02$ dB for each additional bit in the \glspl{adc}. In addition, the dynamic range of a $b$-bit \gls{adc} is defined as the ratio of the largest to the smallest change in the output signal, i.e., $\Delta(2^b-1)/\Delta$, and can be approximated as $6.02b$ dB \cite{baker2008cmos}.

It is observed from \eqref{eq:step_size} that the quantization step size increases with stronger interference. This rise in the quantization step increases the variance of the quantization noise, reducing SQNR in  \eqref{eq:sqnr}, and consequently, communication performance will be significantly degraded in the presence of strong interference.

\subsection{Hypothesis Testing}
The hypothesis testing problem, which accounts for the effects of quantization noise on detecting \gls{bde} information bits, is formulated using two hypotheses: $\mH_{0}$ and $\mH_{1}$. These hypotheses correspond to the cases where the transmitted bit is $``0"$ and $``1"$, respectively, and given by
\begin{equation} 
		\mH_{i}:  \by_j = Q ( \bHdl \bx +\gamma^i_j \bHbl \bx+ \bw_j ),
\end{equation}
where $i \in \{0,1\}$. The function $Q(\cdot)$ is applied to the real and the imaginary parts of the input signal separately. 
The vector $\by_j \in \complexset{\nr}{1}$ is the received signal at the readers in the $j$-th time slot $(j = 1,\dotsc, J)$ and $\bx$ is the beamforming vector showing the transmitted signal from all the \ces. 
The beamformer $\bx$
is designed to enhance the \gls{bc} performance in the presence of \gls{dli}, and its design is detailed in Sections \ref{sec:problems} and \ref{sec:all_bf_sol}.
The scalar $\gamma_j^i$ shows the reflection coefficient of the \gls{bde} in time slot $j$ under hypothesis $\mH_i$.
For simplicity, we assume that $\abs{\gamma_j^i}^2 = \delta, \forall i,j$, where $\delta$ is a constant. 
The vector $\bw_j \in \complexset{\nr}{1}$ is the additive Gaussian noise and all its elements are \gls{iid} $\cgaus{0}{1}$.

Considering quantization, the hypotheses become 
\begin{equation} \label{eq:hypothesisTesting}
\mH_{i}:  \by_j =\bHdl \bx +\gamma^i_j \bHbl  \bx+ \bw_j + \bn_j,
\end{equation}
where $\bn_j$ stands for the quantization noise after \gls{adc}, and its elements follow independent Gaussian distribution. The $r-$th element of $\bn_j$, i.e., $n_{j,r}$, is modeled as $\cgaus{0}{\sigma_{j,r}^2}$ with the variance given by
\begin{equation}\label{eq:variance_of_D}
\resizebox{0.9\linewidth}{!}{$
	\sigma_{j,r}^2 = \left(\ex{\abs{\bh_{\text{DL},r}^\trp \bx}^2} + \ex{\delta \abs{ \bh_{\text{BL},r}^\trp  \bx}^2} + 1\right) / (2^{2b_r} \times 3),
    $}
\end{equation}
where $r=1,\dotsc, \nr$, and $\bh_{\text{BL},r}^\trp \in \complexset{1}{\nc}$ and $\bh_{\text{DL},r}^\trp \in \complexset{1}{\nc}$ are the $r$-th rows of $\bH_{\text{BL}}$ and $\bH_{\text{DL}}$, respectively. The variable $b_r$ denotes the resolution of the \gls{adc} associated with the receive antenna corresponding to the $r$-th row of $\bH_{\text{BL}}$. The covariance matrix of $\bw_j + \bn_j$ is defined as $\nr \times \nr$ diagonal matrix $\bD_j = \diag{\sigma_{j,1}^2+1, \dotsc, \sigma_{j,\nr}^2+1}$.

\subsection{Perfect CSI}
In this subsection, we assume perfect \gls{csi} and derive the optimal \gls{map} detector and its probability of error. \footnote{While the detector structure is conventional, incorporating quantization noise into the analysis and deriving the probability of error are crucial for evaluating the performance under realistic low-resolution conditions.}
Assuming $P(\mH_{0})=P(\mH_{1})=1/2$, the optimal \gls{map} detector is
\begin{equation}\label{eq:lr}
	\frac{ 
		\prod_{j} p\left(\by_j \mid \mH_{1}\right)}
	{\prod_{j} p\left(\by_j \mid \mH_{0}\right)} \underset{\mH_{0}}{\overset{\mH_{1}}{\gtrless}} 1,
\end{equation}
where $p\left(\by_j \mid \mH_{i}\right)$ is the \gls{pdf} of the received signal under $\mH_{i}$, and it is expressed as
\begin{equation} \label{eq:pdf}
	\allowdisplaybreaks
	\begin{split}
		&p\left(\by_j \mid \mH_{i}\right) = \frac{1}{\pi^{\nr} \detr{\bD_j}} \\ &\quad \exp \left[-\norm{ \bD_j^{-1/2}\left( \by_j-\bHdl \bx - \gamma_j^i \bHbl \bx\right)}^2\right].
	\end{split}
\end{equation}

We first define $\by_j^\prime = \by_j-\bHdl\bx$, and then substitute the \glspl{pdf} into \eqref{eq:lr}. Then, by taking the logarithm of both sides in \eqref{eq:lr}, the detector can be expressed as
\begin{equation}\label{eq:llr_PCSI}	
\begin{split}
	&\sum_{j} \bigg\{\norm{\bD_j^{-1/2}\left(\by_j^\prime - \gamma_j^0 \bHbl \bx\right)}^2 - \\ &\quad \norm{\bD_j^{-1/2}\left(\by_j^\prime - \gamma_j^1 \bHbl \bx\right)}^2\bigg\} \underset{\mH_{0}}{\overset{\mH_{1}}{\gtrless}} 0.
    \end{split}
\end{equation}
Using the identity $\norm{\ba - \bb}^2=\norm{\ba}^2+\norm{\bb}^2-2\re{\ba^\herm \bb}$ and  $\abs{\gamma_j^i}^2 = \delta, \forall i,j$,
the detector can be re-expressed as follows
\begin{equation}\label{eq:final_detector}
	\mathsf{LLR} = \sum_{j}\re{(\gamma_j^1 - \gamma_j^0){\by^\prime_j}^\herm \bD_j^{-1} \bHbl \bx}  \underset{\mH_{0}}
	{\overset{\mH_{1}}{\gtrless}}
	0,
\end{equation}
where $\mathsf{LLR}$ is the test statistic.
Under $\mH_{i}$, the test statistic is
\begin{equation}
	\begin{split}
	\mathsf{LLR} &=  \sum_{j}\re{(\gamma_j^1 - \gamma_j^0)(\gamma_j^i)^*}\norm{\bD_j^{-1/2} \bHbl \bx}^2 \\ &\quad + \re{(\gamma_j^1 - \gamma_j^0) (\bw_j + \bn_j)^\herm \bD_j^{-1} \bHbl \bx}.
	\end{split}	
\end{equation} 
The distribution of the test statistic under $\mH_{i}$ is
\begin{equation} \label{eq:dist_under_H1}
\begin{split}
    \mathsf{LLR} &\sim \mathcal{N}\bigg(\sum_{j}\re{(\gamma_j^1 - \gamma_j^0)(\gamma_j^i)^*}\norm{\bD_j^{-1/2} \bHbl \bx}^2, \\ &\quad
\sum_{j}\frac{\abs{\gamma_j^1 - \gamma_j^0}^2}{2}  \norm{\bD_j^{-1/2} \bH_{\text{BL}} \bx}^2\bigg).
    \end{split}
\end{equation}

The probability of error is
$
	P_\text{e} = P(\mH_{0}) P(\mH_{1} | \mH_{0}) + P(\mH_{1}) P(\mH_{0} | \mH_{1}),
$
and it is calculated as 
\footnote{Note that the i.i.d. quantization noise assumption leads to approximate $P_\text{e}$ values in high-\gls{snr} region.}
\begin{equation} \label{eq:P_e}
	P_\text{e} \stackrel{(a)}{=} Q\left(\norm{\bD_j^{-1/2} \bH_{\text{BL}} \bx} \sqrt{\sum_{j}\frac{\abs{\gamma_j^1 - \gamma_j^0}^2}{2}}\right),
\end{equation}
where $Q(x)=\frac{1}{\sqrt{2 \pi}} \int_x^{\infty} \exp \left(-\frac{u^2}{2}\right) d u$. In $(a)$, we use the equality $\re{-(\gamma_j^1-\gamma_j^0)(\gamma_j^0)^*} = \abs{\gamma_j^1 - \gamma_j^0}^2 / 2$. 
For infinite resolution \glspl{adc}, $\bD_j$ is an identity matrix, and the probability of error is defined as $P_\text{e}^\text{inf}$.
It can be showed $P_\text{e}^\text{inf} \leq P_\text{e}$ because each diagonal element of $\bD_j^{-1/2}$ is less than or equal to $1$ in~$P_\text{e}$.

\subsection{Imperfect CSI} 
When there is no perfect \gls{csi}, the estimated channel coefficients are used in a mismatch detector which can be expressed similarly to \eqref{eq:final_detector} as 
\begin{equation}\label{eq:ua_detector}
	\sum\nolimits_{j}\re{(\gamma_j^1 - \gamma_j^0) \hat{\by}_j^{\prime\herm} \hat{\bD}_j^{-1} \hat{\bH}_\text{BL} \bx}  \underset{\mH_{0}}
	{\overset{\mH_{1}}{\gtrless}} 0,
\end{equation}
where $\hat{\by}_j^{\prime} = \by_j-\hat{\bH}_\text{DL}\bx$, $\hat{\bH}_\text{BL} = \hat{\bh}_\text{R} \hat{\bh}_\text{C}^\trp$ and $\hat{\bD}_j$ is created using the estimated \gls{csi} instead of perfect \gls{csi} in \eqref{eq:variance_of_D}. 

Under $\mH_{i}$, the test statistic is given by
\begin{equation}
	\begin{split}
	\mathsf{LLR}' &= \sum\nolimits_{j} \operatorname{Re}\Big\{(\gamma_j^1 - \gamma_j^0) \\ &\quad(\gamma_j^i \bHbl \bx + \bEdl \bx + \bw_j + \bn_j)^\herm \hat{\bD}_j^{-1} \hat{\bH}_\text{BL} \bx\Big\},
	\end{split}	
\end{equation}
where $\bEdl$ models the channel estimation error for $\bHdl$, and each element of it is assumed to have \gls{iid} $\cgaus{0}{\sigma_\text{DL}^2}$. Therefore, each element of the vector $\bEdl \bx$ follows \gls{iid} $\cgaus{0}{\norm{\bx}^2 \sigma_\text{DL}^2}$ and $\norm{\bx}^2 \sigma^2_{\text{DL}} \eye_{N_\text{R}}$ is the covariance matrix of $\bEdl \bx$. The mean under $\mH_{i}$ and the variance of the test statistic are calculated, respectively, as follows
\begin{equation}
    \begin{split}
    \mu_{\mathsf{LLR}' \mid \mH_{i}}  &= \sum\nolimits_{j}\operatorname{Re}\Big\{(\gamma_j^1 - \gamma_j^0)(\gamma_j^i \bHbl \bx)^\herm \hat{\bD}_j^{-1} \bHblHat \bx\Big\}, \\
   \sigma^2_{\mathsf{LLR}'} &= \frac{1}{2} \sum\nolimits_{j} \abs{\gamma_j^1 - \gamma_j^0}^2 \norm{\bC_j \bHblHat \bx}^2,
    \end{split}
\end{equation}
where $\bC_j = \left(\hat{\bD}_j^{-1}(\bD_j + \eye_{\nr} \norm{\bx}^2 \sigma^2_{\text{DL}} )\hat{\bD}_j^{-1}\right)^{1/2} \in \realset{\nr}{\nr}$ is a diagonal matrix.
As a result, the distribution of the test statistic under $\mH_{i}$ is 
$
\mathsf{LLR}'~$ $\sim~\rgaus{\mu_{\mathsf{LLR}' \mid \mH_{i}}}{\sigma^2_{\mathsf{LLR}'}}.
$
The probability of error is
\begin{equation} \label{eq:Pe_ICSI}
    P_\text{e} = \frac{1}{2} Q \left( -\frac{ \mu_{\mathsf{LLR}' \mid \mH_{0}} }{  \sigma_{\mathsf{LLR}'} } \right) + \frac{1}{2} Q \left( \frac{ \mu_{\mathsf{LLR}' \mid \mH_{1}} }{  \sigma_{\mathsf{LLR}'} } \right).
\end{equation} 	

Under perfect CSI, as shown in \eqref{eq:llr_PCSI}, the detector multiplies the received signal by $\bD_j^{-1/2}$, weighting the contribution of each receive antenna according to the total noise variance. The same weighting also appears in \eqref{eq:P_e}. This pre-whitening operation ensures that
antennas with lower noise variance have a stronger influence on the detection outcome.
However, in the case of imperfect CSI, the estimated $\hat{\bD}_j^{-1}$ causes suboptimal noise equalization, as seen in \eqref{eq:ua_detector}, and this effect also appears in \eqref{eq:Pe_ICSI}.
Consequently, $P_\text{e}$ degrades under imperfect \gls{csi} due to both suboptimal pre-whitening and channel estimation errors.

\subsection{Multiple \glspl{bde} Case}
In the case of multiple \gls{bde}, the hypotheses for the $k$-th \gls{bde} are defined as
\begin{equation}
   \mH_{i}^k:  \by_j =\bHdl \bx +\gamma^i_{j,k} \bH_{\text{BL}}^k  \bx+ \sum_{k' \neq k}\gamma^i_{j,k'} \bH_{\text{BL}}^{k'}  \bx+ \bw_j + \bn_j, 
\end{equation}
where each \gls{bde} uses an orthogonal reflection coefficient sequence, and $J \geq K$. Therefore, by using $\gamma_{j,k}^i$ in place of $\gamma_j^i$, one can show that the detectors in
\eqref{eq:final_detector} and \eqref{eq:ua_detector}, as well as the error probability expressions in \eqref{eq:P_e} and \eqref{eq:Pe_ICSI} remain valid for the $k$-th \gls{bde} under the assumption that $\abs{\gamma_{j,k}^i}^2 = \delta_k, \forall i,j$, where $\delta_k$ is a constant.

\section{Problem Definitions}\label{sec:problems}
As seen in~\eqref{eq:P_e}, increasing $\norm{\bD_j^{-1/2} \bH_{\text{BL}} \bx}$ reduces $P_\text{e}$. 
While directly maximizing this term seems straightforward, it is highly sensitive to inaccuracies in the quantization noise model. Moreover, $\bD_j$ depends on both direct and backscatter channels, which may not be reliably known in practice. In addition, maximizing $\norm{\bD_j^{-1/2} \bH_{\text{BL}} \bx}$ is considerably more complicated than maximizing $\norm{\bH_{\text{BL}} \bx}$ due to the dependence of $\bD_j$ on $\bx$.
To enhance robustness under such imperfections, we instead maximize $\norm{\bHbl \bx}$, and introduce an \gls{sir} constraint to suppress \gls{dli} and indirectly regulate $\bD_j$.\footnote{Nonetheless, it is possible to maximize $ \norm{\bD_j^{-1/2} \bHbl \bx}$, for example, using the fractional
optimization method from \cite{shen2018fractional}. The details are provided in the Appendix. The results show that the performance gain achieved by optimizing $\norm{\bD_j^{-1/2} \bHbl \bx}^2$
instead of $\norm{\bHbl \bx}^2$ is minuscule.}

To maximize $\norm{\bHbl \bx}$ in \gls{bibc} over the distributed \gls{mimo} setup, we consider the joint \gls{ap} role selection and \gls{bf} vector design. This objective is formulated as an optimization problem under seven different scenarios.
The first two scenarios do not impose any constraints on \gls{dli}, while the remaining scenarios include constraints to mitigate \gls{dli}. The details of each scenario are outlined in this section, and the proposed solutions are discussed in the following sections.

Note that AP selection directly affects the objective $\|\bH_{\text{BL}} \bx\|^2 = \|\bhr\|^2 |\bhc^\trp \bx|^2$ by determining the channels $\bhc$ and $\bhr$, thus influencing the backscattered signal power.
In DLI-constrained problems, where $\bx$ lies in the null space of the direct-link channel, AP selection also shapes this null space, and enables a trade-off between suppressing \gls{dli} and maximizing $\norm{\bHbl \bx}^2$.
Hence, optimizing AP selection is essential not only for maximizing signal power but also for enabling effective interference control.

\subsection{Maximizing the Received Backscatter Energy}
In this scenario, we does not consider the \gls{dli} constraint, thus the optimization problem can be formulated as
\begin{subequations} \label{eq:op_mrt_1}
	\begin{alignat}{2}
		\Pbf: \quad & \underset{\bx \in \complexset{\nc}{1}, \mathcal{S}}{\text{maximize}} \quad
		& & \ \norm{\bHbl  \bx}^2 \\
		& \text{subject to (s.t.)} \label{con:power_con_mrt}
		& &\  \norm{\bx}^2 \leq P_\text{max}, \\
        & & & 
        \ \apref\in \mS_\text{R}.
        \label{con:ref_in_reader}
	\end{alignat}
\end{subequations}
The constraint \eqref{con:power_con_mrt} limits the total transmitted energy, and \eqref{con:ref_in_reader} guarantees that the $\apref$ is in the reader set. 
Although the problem disregards \gls{dli}, except for the constraint \eqref{con:ref_in_reader} helping to operate under \gls{dli}, it is expected that interference reduces compared to the case of omnidirectional radiation due to the focused energy on the \gls{bc} link. 

On the other hand, in practical systems, each antenna is equipped with its power amplifier. The transmitted energy per antenna can also be constrained to use power resources efficiently. In this case, a new formulation, $\Pbf^\prime$, is defined by replacing constraint \eqref{con:power_con_mrt} with $\abs{x_c}^2 \leq P_\text{max}, c=1,\dotsc,\nc$,
where $x_c$ is the $c$-th element of the vector~$\bx$.

\subsection{\gls{dli} Cancellation}
In the case of limitations on the dynamic range of the \glspl{adc} in the reader circuitry, we should consider \gls{sir} in the reader when designing the transmit beamforming vector to maximize the error performance in \gls{bibc}. This is because the dynamic range of the received signal, and consequently the quantization error, increases with the decreasing $\text{SIR}_r$ in the $r-$th received antenna, which is  expressed as 
\begin{equation}
	\text{SIR}_r = \frac{1}{\eta_r} = \abs{\bh_{\text{BL},r}^\trp \bx}^2 \Big/ \abs{\bh_{\text{DL},r}^\trp \bx}^2,
\end{equation}
where $\eta_r$ is the dynamic range of the received signal in $r-$th received antenna.

Therefore, we formulate the problem $\Pdli$ as the maximization of the received backscattered energy subject to a constraint on the \gls{dli}, and it can be expressed as
\begin{equation} \label{eq:op2}
	\begin{aligned}
		 \Pdli: \quad & \underset{\bx \in \complexset{\nc}{1}, \mathcal{S}}{\text{maximize}}
		& & \norm{\bHbl  \bx}^2 \\
		& \multicolumn{1}{c}{\text{s.t.}}
		& & \eqref{con:power_con_mrt}, \eqref{con:ref_in_reader}, \frac{\abs{\bh_{\text{DL},r}^\trp \bx}^2}{\abs{\bh_{\text{BL},r}^\trp \bx}^2} \leq \alpha,\forall r/r_\refer,
	\end{aligned}
\end{equation}
where $r_\refer$ corresponds to the antenna indices of the $\apref$. The \gls{sir} at the antennas of the reference AP is excluded from the constraint, as these antennas are equipped with high-resolution \glspl{adc}.
The required dynamic range of the \glspl{adc} in the reader is proportional to $\alpha$. A reduction of $6$ dB in $\alpha$ approximately corresponds to $6$ dB increase in SQNR, given in~\eqref{eq:sqnr}, particularly in scenarios with strong \gls{dli}. Therefore, for each $6$ dB reduction in $\alpha$, the number of bits required in \glspl{adc} can be reduced by one while maintaining the same SQNR.

The choice of $\alpha$ depends on use-case requirements and reader \gls{adc} resolution. Increasing $\alpha$ increases received backscatter energy but also raises \gls{dli}. For large $\alpha$, $ \Pdli$ converges to~$\Pbf$.

Alternatively, one can also limit the transmitted energy per antenna. In this case, the new problem $\Pdli^\prime$ is defined by replacing constraint \eqref{con:power_con_mrt} in $\Pdli$ with $\abs{x_c}^2 \leq P_\text{max}, c=1,\dotsc,\nc$.

\subsection{Complete \gls{dli} Cancellation}
In this scenario, it is considered that \gls{dli} is completely mitigated, i.e. $\alpha = 0$, thus the \gls{dli} constraint in~\eqref{eq:op2} can be written as $\bh_{\text{DL},r}^\trp \bx =0, \forall r/r_\refer$.
This constraint can be simplified as $\bH_{\text{DL}}' \bx =0$, which means that $\bx$ lies in the nullspace of $\bH_{\text{DL}}'$, i.e., $\bx \in N(\bH_{\text{DL}}')$. The matrix $\bH_{\text{DL}}' \in \complexset{\nr - M_\refer}{\nc}$ is obtained by removing the $M_\refer$ rows of $\bH_{\text{DL}}$ 
corresponding to the channel between the \glspl{ce} and the $\apref$.
Therefore, the optimization problem can be formulated as
\begin{equation} \label{eq:op3}
	\begin{aligned}
		 \Palpha: \quad & \underset{\bx \in \complexset{\nc}{1}, \mathcal{S}}{\text{maximize}} \quad
		& & \norm{\bHbl  \bx}^2 \\
		& \multicolumn{1}{c}{\text{s.t.}}
		& &  \eqref{con:power_con_mrt}, \eqref{con:ref_in_reader}, \bx \in N(\bH_{\text{DL}}').
	\end{aligned}
\end{equation}
Alternatively, one can also limit the transmitted energy per antenna. In this case, the new problem $ \Palpha^\prime$ is defined by replacing constraint \eqref{con:power_con_mrt}  with $\abs{x_c}^2 \leq P_\text{max}, \forall c$.

\subsection{Multiple \glspl{bde} Case}
For the multiple \glspl{bde} case, we first define the \gls{sinr} for the $k$-th \gls{bde} as 
\begin{equation}
    \text{SINR}_k = \frac{\delta_k \abs{\bu_k^\herm  \bHblk  \bx}^2}{\abs{\bu_k^\herm \bHdl  \bx}^2 +  \sum_{k' \neq k} \delta_{k'} \abs{\bu_k^\herm \bHblkp  \bx}^2 + \norm{\bu_k }^2},
\end{equation}
where $\bHblk = \bhrk (\bhck)^\trp$, $\bu_k = \bhrk$, and $\abs{\gamma^i_{j,k}}^2 = \delta_k, \forall i,j$.
The problem is defined as 
\begin{equation} 
	\begin{aligned}
		 \mP_\text{multi}: \quad & \underset{\bx \in \complexset{\nc}{1}, \mS}{\text{maximize }}  \underset{k}{\text{ min}}\quad
		& & \text{SINR}_k \\
		& \multicolumn{1}{c}{\text{s.t.}}
		& & \eqref{con:power_con_mrt}, \bx \in N(\bH_{\text{DL}}').
	\end{aligned}
\end{equation}
Similar to the other problems, we exclude quantization noise from the \gls{sinr} definition to improve the robustness of $\mP_\text{multi}$ against modeling inaccuracies. We instead introduce a \gls{dli} constraint in the problem formulation.

Table \ref{tab:problems} summarizes the optimization problems by comparing the transmitted energy and the \gls{dli} constraints. Additionally, it provides the corresponding equation numbers for the solutions for \gls{bf} design for the given \gls{ap} partitioning. The detailed solutions for both \gls{bf} design and \gls{ap} selection, for all scenarios, are presented in Sections \ref{sec:all_bf_sol} and \ref{sec:ap_selection}, respectively.

\begin{table}[tbp]
	\centering
	\caption{Summary of the problems}
	\label{tab:problems}
	\resizebox{0.48\textwidth}{!}{\begin{tabular}{|c|c|c|c|}
			\hline
			\rowcolor[HTML]{C0C0C0} 
			\textbf{Problem} & \textbf{Energy Constraint} & \textbf{DLI Constraint} & \textbf{BF Design}  \\ \hline
			$\Pbf$  & \multirow{4}{*}{$\norm{\bx}^2 \leq P_\text{max}$}  & $-$ & \eqref{eq:sol_mrt} \\ \cline{1-1} \cline{3-4} 
			$\Pdli$    &  & $\eta_r \leq \alpha , \forall r/r_\refer$ & \eqref{eq:sol_dli}  \\ \cline{1-1} \cline{3-4}
			$\Palpha$          &   & $\bHdl' \bx = 0$  & \eqref{eq:sol_bf_3}  \\ \cline{1-1} \cline{3-4}
            $\mP_\text{multi}$  &   & $\bHdl' \bx = 0$  & Section \ref{sec:P_multi}  \\ \hline
			$\Pbf^\prime$    & \multirow{3}{*}{$\abs{x_c}^2 \leq P_\text{max}, \forall c$} & $-$  & \eqref{eq:sol_for_mrt2}  \\ \cline{1-1} \cline{3-4} 
			$\Pdli^\prime$       &  & $\eta_r \leq \alpha , \forall r/r_\refer$ &  Section \ref{sec:bf_for_op2} \\ \cline{1-1} \cline{3-4} 
			$\Palpha^\prime$   &   &  $\bHdl' \bx = 0$   & Section \ref{sec:sol_for_p_a0_prime} \\ \hline
	\end{tabular}}
\end{table}

Note that, to isolate and focus on one of the dominant performance bottlenecks of \gls{bc}, which is the quantization error due to the \gls{dli} in the reader, we assume ideal (high-resolution) \glspl{dac}. This assumption maintains analytical tractability and allows the transmit vector $\bx$ to remain unconstrained. Unlike ADCs, DACs do not suffer from DLI, and precoding designs with low‐resolution DACs are well-studied in the literature \cite{mezghani2022massive, jacobsson2017quantized}. It has been shown that low-resolution DACs can approach the performance of ideal \glspl{dac} \cite{jacobsson2017quantized}. In addition, phase-only BF design is more beneficial in the low-resolution DAC scenario, and $\Pdli'$, whose optimal \gls{bf} solution is approximately phase-only, is well suited for such scenarios.

\section{Proposed Beamforming Designs} \label{sec:all_bf_sol}
In this section, we design the beamforming vector to maximize the received backscattered energy for a given set $\mS$. 

\subsection{Beamforming Design for $\mathcal{P}_{\text{\textnormal{BF}}}$}
The problem $\Pbf$ for the given set $\mS$ can be written as 
\begin{equation} \label{eq:op_mrt}
\mathcal{P}_{\text{BF},1}: \underset{\bx \in \complexset{\nc}{1}}{\text{maximize }}
		 \bx^\herm \bHbl^\herm \bHbl \bx, \text{ s.t. }
		 \norm{\bx}^2 \leq P_\text{max}. 
\end{equation} 
This problem is known as the Rayleigh quotient, and the optimal $\bx$ is the right singular vector of $\bHbl$, where $\rank{\bHbl}=1$.
The optimal solution is 
\begin{equation}\label{eq:sol_mrt}
	\bxbf = \sqrt{\Pmax}\bv_\text{BL} = \sqrt{\Pmax} e^{j\theta}  \bhc^*/\norm{\bhc},
\end{equation}
where $\bv_\text{BL} \in \complexset{\nc}{1}$ is the
right singular vector with the unit norm and $\bxbf$ corresponds to the \gls{mrt}.
The term $e^{j\theta}$ represents a common phase rotation applied to all transmit antennas. The phase $\theta \in [0,2\pi)$ does not affect the constructive summation of the transmitted signals at the reader, as the relative phase alignment between antennas remains unchanged. The solution is therefore not unique, and any constant phase shift $\theta$ yields the same performance.

\subsection{Beamforming Design for $\mathcal{P}'_{\text{\textnormal{BF}}}$}
The problem $\Pbfp$ for the given set $\mS$ can be written as 
\begin{equation} \label{eq:op1'}
\mathcal{P}_{\text{BF},1}^\prime: \underset{\bx \in \complexset{\nc}{1}}{\text{maximize }}
\norm{\mathbf{\sigma}_\text{BL} \bu_\text{BL} \bv_\text{BL}^\herm  \bx}^2, \text{ s.t. }
\abs{x_c}^2 \leq P_\text{max}, \forall c, 
\end{equation} 
where the \gls{svd} of $\bHbl = \mathbf{\sigma}_\text{BL} \bu_\text{BL} \bv_\text{BL}^\herm$ and $\bu_\text{BL} \in \complexset{\nr}{1}$ has unit norm. The problem can be equivalently expressed as
\begin{equation} 
		\mathcal{P}_{\text{BF},1}^\prime:  \underset{\bx \in \complexset{\nc}{1}}{\text{maximize }}
		 \abs{ \sum_{c = 1}^{\nc} v_{\text{BL},c}^* x_c}^2, 
	\text{ s.t. }
		 \abs{x_c}^2 \leq P_\text{max}, \forall c,
\end{equation}
where $v_{\text{BL},c}$ is the $c$-th element of $\bv_{\text{BL}}$, and the objective function is maximized for
\begin{equation} \label{eq:sol_for_mrt2}
	x_c = \sqrt{\Pmax} \frac{v_{\text{BL},c}}{\abs{v_{\text{BL},c}}} = \sqrt{\Pmax} e^{j\theta}  \frac{h_{\text{C}, c}^*}{\abs{h_{\text{C},c}}},
\end{equation}
where $h_{\text{C}, c}$ and $x_c$ are the $c$-th element of $\bh_{\text{C}}$ and the optimal \gls{bf} vector $\bx_{\text{BF}^\prime}$, respectively. 

\subsection{Beamforming Design for $\mathcal{P}_{\text{\textnormal{DLI}}}$ and $\mathcal{P}'_{\text{\textnormal{DLI}}}$} \label{sec:bf_for_op2}
Using the equality $\norm{\bA}^2 = \tr{\bA \bA^\herm}$ and the cyclic property of the trace operator, the problem $ \Pdli$ for the given set $\mS$ can be written as 
\begin{equation} \label{eq:op2_real_trace}
	\begin{aligned}
		\mP_{\text{DLI},1}: \quad & \underset{\bx \in \complexset{\nc}{1}}{\text{maximize}}
		& & \tr{\bH_\text{BL}^\herm \bH_\text{BL} \bx {\bx}^\herm} \\
		& \multicolumn{1}{c}{\text{s.t.}}
		& & \frac{\tr{\bh_{\text{DL},r}^* \bh_{\text{DL},r}^\trp \bx {\bx}^\herm}}{\tr{\bh_{\text{BL},r}^* \bh_{\text{BL},r}^\trp \bx {\bx}^\herm}} \leq \alpha, \forall r/r_\refer, \\ 
		& & & \tr{\bx  {\bx}^\herm} \leq P_\text{max}. \\
	\end{aligned}
\end{equation}
The solution is obtained through two steps as described below.

\subsubsection{Step 1 - Semidefinite Relaxation}
Let us define  $\bG_\text{BL}$$=$$\bH_\text{BL}^\herm \bH_\text{BL} $, $\bG_\text{BL}^r$$=$$\bh_{\text{BL},r}^* \bh_{\text{BL},r}^\trp$,
$\bG_\text{DL}^r$$=$$\bh_{\text{DL},r}^* \bh_{\text{DL},r}^\trp$,  and $\bX$$=$$\bx {\bx}^\herm$. Then, the problem in~\eqref{eq:op2_real_trace} is
\begin{equation} \label{eq:op2_real_trace2}
	\begin{aligned}
		\quad & \underset{\bX \in \complexset{\nc}{\nc}}{\text{maximize}}
		& & \tr{\bG_\text{BL} \bX} \\
		& \multicolumn{1}{c}{\text{s.t.}}
		& & \tr{(\bG_\text{DL}^r-\alpha \bG_\text{BL}^r) \bX} \leq 0,  \forall r/r_\refer, \\
		& & & \tr{\bX} \leq P_\text{max}, \bX \succeq 0, \rank{\bX} = 1.
	\end{aligned}
\end{equation}

The problem in \eqref{eq:op2_real_trace2} is a non-convex optimization problem due to the constraint $\rank{\bX} = 1$. 
Thus, we can apply \gls{sdr} by dropping the rank-$1$ constraint to convert the problem into a convex optimization problem, $\mathcal{P}_{\text{SDR}}$,  with a globally optimal solution.

\subsubsection{Step 2 - Optimization Toolbox for the Solution} 
The problem $\mathcal{P}_{\text{SDR}}$ can be solved using the interior point method and \gls{cvx}. The global optimal solution of the problem $\mathcal{P}_{\text{SDR}}$ is denoted as $\bX_\text{opt}$. 

The best rank-$1$ approximation of $\bX_\text{opt}$ is given by $\lambda_\text{opt} \bq_\text{opt} \bq_\text{opt}^\herm$, where $\lambda_\text{opt} \in \mathbb{C}$ is the largest eigenvalue of $\bX_\text{opt}$, and $\bq_\text{opt} \in \complexset{\nc}{1}$ is the corresponding eigenvector and $\norm{\bq_\text{opt}}$$=$$1$ \cite{eckart1936approximation}.
Then, the solution to the problem $\mP_{\text{DLI},1}$ 
is the scaled version of the dominant eigenvector of $\bX_\text{opt}$ given by
\begin{equation}\label{eq:sol_dli}
\bx_\text{DLI} = \sqrt{P_\text{max}}\bq_\text{opt}.
\end{equation}

\textbf{\gls{bf} design for} $\Pdli^\prime$:
For the scenario with per antenna constraint, $\Pdli^\prime$ with the given set $\mS$, we can also follow the aforementioned steps. Thus, the relaxed problem for $\Pdli^\prime$ can be defined by just replacing the constraint $\tr{\bX} \leq P_\text{max}$ in ~\eqref{eq:op2_real_trace2} by $[\bX]_{c,c} \leq P_\text{max}, \forall c$, which limits the transmitted energy per transmit antenna. The solution for $\Pdli^\prime$ is called $\bx_{\text{DLI}^\prime}$ which can be calculated similar to \eqref{eq:sol_dli}. 

\subsection{Beamforming Design for $\Palpha$} \label{sec:bf_for_op3}
The problem $\Palpha$ for the given set $\mS$ can be written as 
\begin{equation}
\Palphaf: \underset{\bx \in \complexset{\nc}{1}}{\text{maximize }}
		 \norm{\bHbl  \bx}^2,
		 \text{ s.t. } \eqref{con:power_con_mrt}, \bx  \in N(\bH_{\text{DL}}').
\end{equation} 
The orthonormal basis for the nullspace of $\bH_{\text{DL}}'$ 
is created using the $\nc - r_\text{DL}$ right singular vectors of $\bH_{\text{DL}}'$ corresponding to its zero singular values. This basis is denoted by
$\bZ_{\text{DL}} \in \complexset{\nc}{\nc-r_\text{DL}}$ where $\bZ_{\text{DL}}^\herm \bZ_{\text{DL}} = \eye_{\nc-r_\text{DL}}$ and $r_\text{DL} = \rank{\bHdl'} \leq \min\{\nr - M_\refer, \nc\}$.
In $\Palphaf$, the vector $\bx$ can be written as $\bx = \bZ_{\text{DL}} \bb$, where $\bb \in \complexset{\nc-r_\text{DL}}{1}$, 
to remove the constraint $\bx  \in N(\bH_{\text{DL}}')$ as 
\begin{equation} \label{eq:op3_prime}
	 \underset{\bb}{\text{maximize }}
		 \norm{\bHbl  \bZ_{\text{DL}} \bb}^2, 
		\text{ s.t. }
	 \norm{\bZ_{\text{DL}} \bb}^2 = \norm{\bb}^2 \leq P_\text{max}.
\end{equation} 
Similar to the solution of~\eqref{eq:op_mrt}, the solution for the above problem is the right singular vector of $\bHbl  \bZ_{\text{DL}}$. Thus, 
the optimal solution for the problem in \eqref{eq:op3_prime} is 
$
\bb_\text{opt} = \sqrt{P_\text{max}}\bv_\text{Z},
$
where $\bv_\text{Z} \in \complexset{\nc-r_\text{DL}}{1}$ is the  right singular vector.
Then, the optimal solution for $\Palphaf$~is 
\begin{equation} \label{eq:sol_bf_3}
	\bx_{\alpha0} = \sqrt{P_\text{max}} \bZ_{\text{DL}} \bv_\text{Z},
\end{equation}
which lies in the nullspace of $\bH_{\text{DL}}'$ because $\bx_{\alpha0}$ is the combination of the columns of $\bZ_{\text{DL}}$. Note that the optimal solution can be rewritten as
\begin{equation} \label{eq:sol_for_alpha0}
		\bx_{\alpha0}  =\sqrt{P_\text{max}} e^{j\theta}  \bZ_{\text{DL}} \bZ_{\text{DL}}^\herm \bhc^* /\norm{\bZ_{\text{DL}}^\herm \bhc^*}, \exists \theta \in \mathbb{R},
\end{equation}
where $\bZ_{\text{DL}} \bZ_{\text{DL}}^\herm$ is a projection matrix to the nullspace of $\bHdl'$.
The optimal \gls{bf} vector in~\eqref{eq:sol_for_alpha0} is the scaled version of the projection of~\eqref{eq:sol_mrt} to the nullspace of $\bHdl'$.

\subsection{Beamforming Design for $\Palpha^\prime$} 
\label{sec:sol_for_p_a0_prime}
The problem $\Palpha^\prime$ for the given set $\mS$ can be written as
\begin{equation} \label{eq:prob_Pa0_prime}
	\begin{aligned}
		\Palphaf^\prime: \quad & \underset{\bx \in \complexset{\nc}{1}}{\text{maximize}}
		& & \norm{\bHbl  \bx}^2 \\
		& \multicolumn{1}{c}{\text{s.t.}}
		& & \bHdl' \bx = 0, \abs{x_c}^2 \leq P_\text{max}, \forall c. \\
	\end{aligned}
\end{equation} 
This is a non-convex problem due to the maximization of the norm. Using the \gls{svd} of $\bHbl$, the objective function can be reformulated as $\abs{\bv_\text{BL}^\herm  \bx e^{j \theta}}^2$, where $\theta$ is an arbitrary phase rotation. Without loss of optimality, $\theta$ can be chosen such that the objective becomes a real-valued quantity, i.e., $\abs{\bv_\text{BL}^\herm  \bx e^{j \theta}} = \re{\bv_\text{BL}^\herm  \bx}$. Therefore, the problem can be equivalently rewritten as
\begin{equation}  \label{eq:prob_Pa0_prime_convex}
	\begin{aligned}
		& \underset{\bx \in \complexset{\nc}{1}}{\text{maximize}}
		& & \re{\bv_\text{BL}^\herm  \bx} \\
		& \multicolumn{1}{c}{\text{s.t.}}
		& & \bHdl' \bx = 0, \abs{x_c}^2 \leq P_\text{max}, \forall c. \\
	\end{aligned}
\end{equation}
Note that this reformulated problem is convex and equivalent to $\Palphaf^\prime$, and can be efficiently solved using standard convex optimization solvers.

\textbf{Closed form \gls{bf} design for $\Palphaf^\prime$:} We relax the constraint $\abs{x_c}^2 \leq P_\text{max}$ in~\eqref{eq:prob_Pa0_prime} to $\norm{\bx}^2 \leq P_\text{max}$. With this relaxation, the solution is the same with~\eqref{eq:sol_bf_3}. Due to the original constraint $\abs{x_c}^2 \leq P_\text{max}$, we scale the solution, and the final sub-optimal closed-form solution can be given as 
\begin{equation} \label{eq:sol_Pa0_prime_closed}
	\bx_{\alpha0^\prime} = \sqrt{P_\text{max}} \bZ_{\text{DL}} \bv_\text{Z} / q_{\text{max}},
\end{equation}
where $q_{\text{max}}$ represents the maximum absolute value among the elements of $\bZ_{\text{DL}} \bv_\text{Z}$.

\subsection{Beamforming Design for $\mP_\text{multi}$}
\label{sec:P_multi}
We use a bisection method to solve \(\mP_\text{multi}\).
The overall bisection method is summarized in Algorithm \ref{alg:bisection}.

We first define $\bx = \bZ_{\text{DL}} \bb$ as in  $\Palpha$. For the given set $\mS$, the problem $\mP_\text{multi}$ is equivalent to 
\begin{equation} 
	\begin{aligned}
		 \quad & \underset{\bb}{\text{maximize }} \quad
		& & t \\
		& \multicolumn{1}{c}{\text{s.t.}}
		& &  \text{SINR}_k \geq t, \forall k, \\
        & & & \norm{\bb}^2 \leq P_\text{max},
	\end{aligned}
\end{equation}
where $t = \underset{k}{\text{ min }} \text{SINR}_k$. For a given \( t \), we check feasibility via the following problem:
\begin{equation} \label{eq:feasibility_problem}
    \begin{aligned}
        & \text{find} && \bb \\
        & \text{s.t.} 
        & &  \text{SINR}_k \geq t, \forall k, \\
        & & & \norm{\bb}^2 \leq P_\text{max}.
    \end{aligned}
\end{equation}
We define the following matrices
\begin{align*}
\bA_k &= \delta_k \norm{\bhrk}^4\bZ_{\text{DL}}^\herm (\bhck)^* (\bhck)^\trp \bZ_{\text{DL}}, \\
\bB_k &= \bZ_{\text{DL}}^\herm \bHdl^\herm \bhrk (\bhrk)^\herm \bHdl \bZ_{\text{DL}}, \\
\bC_{k} &= \bZ_{\text{DL}}^\herm \sum_{k' \neq k} \left(\delta_{k'} (\bHblkp)^\herm \bhrk (\bhrk)^\herm \bHblkp \right) \bZ_{\text{DL}}.
\end{align*}
Using these matrices, the SINR constraint for the $k$-th \gls{bde} is
\begin{equation} \label{eq:SINR_cons}
    \bb^\herm \left( \bA_k - t (\bB_k + \bC_k) \right) \bb \geq \norm{\bhrk}^2 t, \quad \forall k.
\end{equation}
The feasibility problem in \eqref{eq:feasibility_problem} is solved by applying semidefinite relaxation as in Section \ref{sec:bf_for_op2}. 
\begin{algorithm}[tbp]
    \caption{Bisection Algorithm for $\mP_\text{multi}$}
    \label{alg:bisection}
    \begin{algorithmic}[1]
        \Require $\bA_k, \bB_k, \bC_k$ for all $k$, $P_\text{max}$, $\bZ_\text{DL}$, and $\epsilon$
        \State \textbf{Initialize:} Initialize the bisection bounds: \( t_{\min} \) and \( t_{\max} \).
        \While{$t_{\max} - t_{\min} > \epsilon$}
            \State Set $t \gets \frac{t_{\min} + t_{\max}}{2}$
            \State Solve the feasibility problem in \eqref{eq:feasibility_problem}.
            \If{the problem is feasible}
                \State Update lower bound: $t_{\min} \gets t$
                \State Update beamforming vector: $\bb^*$
            \Else
                \State Update upper bound: $t_{\max} \gets t$
            \EndIf
        \EndWhile
        \Ensure The beamforming vector $\bZ_{\text{DL}} \bb^*$
    \end{algorithmic}
\end{algorithm}

\section{Proposed AP Partitioning Algorithms}\label{sec:ap_selection}

In this section, we propose algorithms for \gls{ap} role selection for all problems outlined in Table \ref{tab:problems}, using the \gls{bf} vectors derived in Section \ref{sec:all_bf_sol}. 
The problem $\Pbf$ is solved using \gls{dp}. The remaining problems are solved using an iterative optimization algorithm.

\subsection{AP Partitioning Algorithm for $\mathcal{P}_{\text{\textnormal{BF}}}$} \label{sec:ap_part_for_pbf}
When we substitute the optimal \gls{bf} vector $\bxbf$ given in \eqref{eq:sol_mrt} to $\Pbf$ given in~\eqref{eq:op_mrt_1}, the constraint $\norm{\bx}^2 \leq P_\text{max}$ holds with equality and $\Pbf$ is reduced solely to the \gls{ap} role selection, and it can be expressed~as
\begin{equation} \label{eq:subsetsum}
		\quad \underset{ \mathcal{S}}{\text{maximize }}
		\norm{\bhr}^2\norm{\bhc}^2, \text{ s.t. }
        \apref\in \mS_\text{R}.
\end{equation}
To maximize the objective function in~\eqref{eq:subsetsum}, $\norm{\bhr}^2$ and $\norm{\bhc}^2$ should ideally be equal to $\sum_{l\in\mL} \norm{\bh_l}^2 / 2$.
As a result, the problem in \eqref{eq:subsetsum} is equivalent to 
\begin{equation} \label{eq:subsetsum2}
		\underset{\mS_\kappa}{\text{maximize }}
		 \sum_{l \in \mS_\kappa}\norm{\bh_l}^2, 
		\text{ s.t.}
		 \sum_{l \in \mS_\kappa}\norm{\bh_l}^2 \leq 
          \sum_{l\in\mL} \norm{\bh_l}^2/2,
\end{equation}
where $\mS_\kappa$ $(\kappa \in \{CE, R\})$ is a subset of $\mathcal{L}$. If the solution set includes $\apref$, $\kappa=\text{R}$ and $\mS_\kappa$ gives the set of readers, otherwise $\kappa=\text{CE}$.
This problem is similar to the subset sum problem, a special case of the Knapsack problem, and can be solved using a greedy algorithm or \gls{dp} \cite{kleinberg2006algorithm}.

Unlike greedy algorithms, \gls{dp} generally converges to optimal solutions.
In \gls{dp}, integers are generally used, thus we reformulate the problem in~\eqref{eq:subsetsum2} by rounding the floats to integers. Let us define integer weights $q_l = \floor{s \norm{\bh_l}^2 + 0.5}$, where $s$ is a scaling factor affecting the precision. Then, the relaxed problem can be formulated as
\begin{equation} \label{eq:subsetsum2_relax}
    \underset{\mS_\kappa}{\text{maximize }}
    \sum_{l \in \mS_\kappa} q_l,
    \text{ s.t. }
    \sum_{l \in \mS_\kappa} q_l \leq \floor{\sum_{l \in \mL} \frac{1}{2} q_l} = Q. 
\end{equation}
For this case, \gls{dp} provides the optimal solution. However, the optimal solution for~\eqref{eq:subsetsum2_relax} may not be optimal for~\eqref{eq:subsetsum2} due to the rounding errors and low precision. 
With the increasing scaling factor, the optimal solutions for~\eqref{eq:subsetsum2_relax} and \eqref{eq:subsetsum2} will approach each other at the cost of increasing the algorithm's complexity.

The \gls{dp} algorithm creates small subproblems and recursively solves these subproblems to reach the final solution. In total, we have $L (Q+1)$ subproblems, one for each element of $\mL$ and $q \in \{0, 1, \dotsc, Q\}$. Let us define a subproblem  for parameters $l'$ and $q$ as follows
\begin{equation} \label{eq:subsetsum2_subproblem}
	\mathcal{P}_{l',q}: \underset{\mS_{l'}}{\text{maximize }}
		 \sum_{l \in \mS_{l'}} q_l, 
		 \text{ s.t. }
	\sum_{l \in \mS_{l'}} q_l \leq q, 	
\end{equation}
where $\mS_{l'} \subseteq \{1,2,\dotsc,l'\}$ is a subset of first $l'$ \gls{ap}, and $OP(l',q)$ denotes the optimal maximum value of the subproblem $\mathcal{P}_{l',q}$.
The initial values are $OP(0,q)=0, \forall q$. Then, we solve the subproblems recursively using the following rule \cite{kleinberg2006algorithm}
\begin{equation} \label{eq:recurrence}
\begin{split}
    &OP(l',q) = \\
    &\begin{cases}
        OP(l'-1,q) & \text{if } q < q_{l'}, \\
        \max\big(OP(l'-1,q), q_{l'} + OP(l'-1,q-q_{l'})\big) & \text{otherwise}.
    \end{cases}
    \end{split}
\end{equation}
The summary of the algorithm is given in Algorithm \ref{alg:dp_alg}. For example, if the weight for $\ap_{l'}$ exceeds the summation constraint $q$, then $\mathcal{P}_{l',q}$ and $\mathcal{P}_{l'-1,q}$ have identical solutions, so $OP(l',q) = OP(l'-1,q)$. Conversely, if the weight of $\ap_{l'}$ is at most $q$, we use the optimal solution calculated previously for $\mathcal{P}_{l'-1,q-q_{l'}}$, aiming to maximize the objective function using the subset of the first $l'-1$ \glspl{ap} with a constraint $q-q_{l'}$. Then, the solution to $\mathcal{P}_{l',q}$ is determined by taking the greater value between $OP(l'-1,q)$ and $q_{l'} + OP(l'-1,q-q_{l'})$. 
This approach reduces complexity by calculating $OP(l',q)$ from $OP(l'-1,q)$ and $OP(l'-1,q-q_{l'})$, achieving a time complexity of $\mO(LQ)$, in contrast to the exponential complexity of $\mO(2^L)$ for an exhaustive search.
More details on \gls{dp} can be found in \cite{kleinberg2006algorithm}.

\begin{algorithm}[tbp]
	\caption{DP for AP Partitioning}\label{alg:dp_alg}
	\begin{algorithmic}[1]
		\Require $\mL, q_k, Q$
		\Ensure $\mS_\kappa$
		\State $OP(0,q)=0, \forall q$
		\For{$l' \gets 1 \text{ to } L$}
		\For{$q \gets 0 \text{ to }  Q$}
		\If{$q_{l'}>q$}
		\State $OP(l',q) = OP(l'-1,q)$
		\State Save the \gls{ap} indices maximizing $OP(l',q)$.
		\Else
        \State $OP(l',q) = \text{max}(OP(l'-1,q),$
        \Statex \hspace{1.8cm} $q_{l'} + OP(l'-1,q-q_{l'}))$
		\State Save the \gls{ap} indices maximizing $OP(l',q)$.
		\EndIf	
		\EndFor
		\EndFor
	\end{algorithmic}
\end{algorithm}

When computing each $OP(l',q)$ value, we store the \aps indices. Finally, the \aps indices for $OP(L,Q)$ are the set of selected \glspl{ce} if $\apref$ is included; otherwise, they correspond to the readers for the problem in \eqref{eq:subsetsum2_relax}. 
In summary, for the problem $\Pbf$, the solution of $OP(L,Q)$ is the \glspl{ce} (or readers) while the remaining \aps are readers (or \glspl{ce}). 

\subsection{AP Partitioning Algorithm for the Remaining Problems}\label{sec:ap_partition}
In this subsection, we give the details of the proposed \gls{ap} role selection algorithm for all problems except $\Pbf$. Searching for all possible combinations of \ces and readers is not a feasible solution because the computational complexity of the exhaustive search increases with increasing number of \aps. Thus, the proposed algorithm uses a coalitional game theory approach to find the set of \ces and readers among the given~\aps.

\subsubsection{Coalition Game}
In this cooperative game, the set of \aps $\mathcal{L}$ are cooperative players that form coalitions. We have two coalitions, sets of \ces $(\mS_\text{CE})$ and readers $(\mS_\text{R})$, and the partition of the given \aps are defined as $\mS=\{\mS_\text{CE}, \mS_\text{R}\}$.
The union of these two sets includes all \aps while the intersection of them is empty as
$\mathcal{L} = \mS_\text{CE} \cup \mS_\text{R}$ and  $\mS_\text{CE} \cap \mS_\text{R} = \varnothing$.

The non-transferable utility function is defined as $U(\mS) =  \norm{\bHbl  \bx}^2$.  In addition, we define the function $C\left(\mS \right) = \max\limits_{r/r_\refer} \left( \abs{\bh_{\text{DL},r}^\trp \bx}^2 / \abs{\bh_{\text{BL},r}^\trp \bx}^2 \right)$, which is used to check whether the constraint on \gls{dli} is satisfied for each reader antenna except the $\apref$ antennas. 
Note that, $U(\mS) =  t$ and $C\left(\mS \right) = \max\limits_{k, r/r_\refer} \left( \abs{(\bh_{\text{DL},r}^k)^\trp \bx}^2 / \abs{(\bh_{\text{BL},r}^k)^\trp \bx}^2 \right)$ for $\mP_\text{multi}$.
If the reader set only includes $\apref$, then $C\left(\mS \right)=0$.
The role of an \gls{ap} depends on the increase/decrease in the utility function and the \gls{dli} constraint. 
After the initial partition, $\ap_l$ will switch its role based on the following operations:
\begin{itemize}
	\item \textit{Switch Operation:} 
	Let us define $\mS_{\kappa}$ as the set of readers (\ces) and $\mS_{\kappa^\prime}$ as the set of \ces (readers) where $\kappa, \kappa' \in \{CE, R\}$. The preference relation $\mS^{\prime} \succ_l \mS$ implies that $\ap_l$ in $\mS_{\kappa}$ prefers to be a member of $\mS_{\kappa^\prime}$, where $\mS^{\prime}$ is the new set of \ces and readers and defined as $\mS^{\prime} = \{\mS_{\kappa}\backslash\{l\}, \mS_{\kappa^\prime} \cup \{l\} \}$. This rule can be written as 
	\begin{equation}\label{eq:operation_1}
		S^{\prime} \succ_l S \Leftrightarrow U\left(\mS_{\kappa}\backslash\{l\}, \mS_{\kappa^\prime} \cup \{l\}\right) > U\left(S \right), C\left(\mS^\prime \right) \leq \alpha,
	\end{equation}
	where $l \in \mL / \refer$. If there is no \gls{dli} constraint, we do not check $C\left(\mS^\prime \right) \leq \alpha$ for switch operation.
	
	\item \textit{Swap Operation:} The swap operation is defined as follows
	\begin{equation}\label{eq:operation_2}
		\mS = \left\{ \mS_{\kappa}, \mS_{\kappa^\prime} \right\} \rightarrow \mS^{\prime} = \left\{
		\mS_{\kappa}\backslash\{l\} \cup \{l^\prime\}, \mS_{\kappa^\prime} \backslash\{l^\prime\} \cup \{l\}
		\right\},
	\end{equation}
    where $l, l' \in \mL / \refer$, and $\ap_l$ swaps its coalition with $\ap_{l'}$ if $U(\mS^\prime)>U(\mS)$ and $C\left(\mS^\prime \right) \leq \alpha$. When there is no \gls{dli} constraint, we only check whether $U(\mS^\prime)>U(\mS)$.
\end{itemize}

Note that for $\mP_\text{multi}$, neither the switch nor the swap operation is performed if the problem in \eqref{eq:feasibility_problem} is infeasible for the given CE and reader sets.

\subsubsection{Proposed Solution Based on Coalition Game Algorithm}
The proposed algorithm consists of four phases:

\begin{algorithm}[tbp]
	\caption{Coalitional Game Algorithm}\label{alg:coalitional_game}
	\begin{algorithmic}[1]
		\Require $\mathcal{L}, \mS_\text{init}, \apref, \alpha, P_\text{max}, \bHbl, \bHdl, \bHdl'$
		\Ensure $\mS, \bx$
        \State $\mS \gets \mS_\text{init}$
		\While{$\text{convergence}=0$}\Comment{We have the solution if convergence is $1$.}
		\State $\mathcal{L}_\text{aux} \gets \mathcal{L}/ \apref$ 
		\State $\mS_\text{prev.} \gets \mS$
		\While{$\abs{\mathcal{L}_\text{aux}} \not= 0$} \Comment{We iterate over all \aps.}
		\State Select a random \gls{ap}, $\ap_l$, from the set $\mathcal{L}_\text{aux}$.
		\If{$\abs{\mS_{\kappa}} \not= 1$}
		\State $\mS_\text{aux} \gets \{\mS_{\kappa}\backslash\{l\}, \mS_{\kappa^\prime} \cup \{l\} \}$.
		\State Calculate $U(\mS_\text{aux}), C(\mS_\text{aux})$ and \gls{bf} vector using the 
        \Statex \hspace{1.35cm} results in Section \ref{sec:all_bf_sol}.
		\If{$\mS_\text{aux} \succ_l \mS$}
		\State $\mS \gets \mS_\text{aux}$
		\State $\bx \gets \bx_\text{aux}$
		\EndIf
		\EndIf	
		\State Delete the $\ap_l$ from the set $\mathcal{L}_\text{aux}$		
		\EndWhile
		\If{$\mS = \mS_\text{prev.}$}
		\State convergence $= 1$.
		\EndIf
		\EndWhile\label{euclidendwhile}
	\end{algorithmic}
\end{algorithm}

\textbf{Phase 1 - Initial Set of \glspl{ce} and Readers:} 
    In $\Pbf^\prime$, there is no \gls{dli} constraint. Therefore, we randomly select \aps as \glspl{ce} and readers and use this initial set as input for Algorithm \ref{alg:coalitional_game}. However, we always put the $\apref$ to the reader set.

    For the remaining problems, which have \gls{dli} constraint, we randomly select an initial set of \glspl{ce} and readers $(\mS_{\text{init}})$, ensuring $\apref$ is in the reader set. 
    Then we verify the \gls{dli} constraint, i.e., $C\left(\mS_\text{init} \right) \leq \alpha$. If it is satisfied, the initial set is provided as input to Algorithm \ref{alg:coalitional_game}. Otherwise, we generate new sets at random until the constraint is met or the number of generated initial sets reaches a predefined threshold, i.e., $\zeta_\text{init}$. If the number of initial sets reaches $\zeta_\text{init}$, the final set is used as input for Algorithm \ref{alg:coalitional_game}, and $U(\mS_\text{init})$ is set to zero if $C\left(\mS_\text{init} \right) > \alpha$.
    
\textbf{Phase 2 - Algorithm \ref{alg:coalitional_game}:} In this phase, Algorithm \ref{alg:coalitional_game} iterates over all \aps except $\apref$, performing the following steps in each iteration:
	\begin{itemize}
		\item \textit{Step 1:} An \gls{ap} is selected randomly.
		\item \textit{Step 2:} If feasible, a switch operation changes the \gls{ap}'s role.
		\item \textit{Step 3:} The \gls{bf} vector is recalculated.
		\item \textit{Step 4:} If the switch operation improves the utility function and meets the \gls{dli} constraint (if applicable), the updated \glspl{ce} and readers set are retained; otherwise, they revert to the original configuration.
	\end{itemize} 
    The steps above are repeated until convergence is achieved.
    In Algorithm \ref{alg:coalitional_game}, let us assume $\ap_l$ is randomly selected in Step 1. Then, $\mS_{\kappa}$ 
    represents the set containing $\ap_l$, while $\mS_{\kappa^\prime}$ denotes the complementary set. If $\ap_l$ is part of the \gls{ce} set, then $\mS_{\kappa}=\mS_{CE}$ and $\mS_{\kappa^\prime} = \mS_{R}$. If $|\mS_{\kappa}|=1$ when $\ap_l$ is selected, the switch operation cannot be applied, and $\mS$ remains unchanged.
    
\textbf{Phase 3 - \gls{dli} Constraint:}  Since the number of possible \gls{ap} partitions is finite, and each switch operation strictly increases the utility function as shown in \eqref{eq:operation_1}, Algorithm \ref{alg:coalitional_game} is guaranteed to converge to a set $\mS^*$ where no feasible switch operation can result in a higher utility. This can be expressed as
    $
        U(\mS^*) \geq U(\mS),
    $
    where $\mS^*$ is the output of Algorithm  \ref{alg:coalitional_game} and 
    no further improvement is possible through any feasible switch operation that leads to a new partition $\mS$.
    After the convergence, Algorithm \ref{alg:coalitional_game} provides the set of \glspl{ce} and readers. 
    
    If the initial set from Phase 1 does not satisfy the \gls{dli} constraint, the output set from Algorithm \ref{alg:coalitional_game} may not satisfy $C\left(\mS \right) \leq \alpha$. If that is the case, Phase 1 and 2 are repeated until either $C\left(\mS \right) \leq \alpha$ is satisfied after Phase 2 or a predetermined threshold for the maximum number of repetitions of the first two phases, i.e., $\zeta_\text{Alg5}$, is reached.

    For $\Pbf^\prime$, Phase 1 and 2 are repeated $\zeta_\text{Alg5}$ times with different initial solutions since there is no \gls{dli} constraint. Then, the best solution is selected.

\textbf{Phase 4 - Algorithm \ref{alg:swap_alg}:} Once Phase 3 is completed, the set $\mS$ is provided as input to Algorithm \ref{alg:swap_alg} which swaps each element of the \gls{ce} set with each element of the reader set except $\apref$. If the utility function improves and \gls{dli} constraint (if present) is satisfied after the swap operation, the changes are kept. The result of Algorithm \ref{alg:swap_alg} is the final set of \ces and readers.

\begin{algorithm}[tbp]
	\caption{Swap Algorithm}\label{alg:swap_alg}
	\begin{algorithmic}[1]
		\Require $\mS, \bx, \apref, \alpha, P_\text{max}, \bHbl, \bHdl$
		\Ensure $\mS, \bx$
		\State Calculate $U(\mS) = \norm{\bH_\text{BL} \bx}^2$.
		\For{$l \gets 1 \text{ to } \abs{\mSce}$}
		\For{$l^\prime \gets 1 \text{ to }  \abs{\mSr}$}
		\State $\mSaux = \Big\{
		\mSce \backslash\{\mSce(l)\} \cup \{\mSr(l^\prime)\},$ 
        \Statex \hspace{1.8cm}$\mSr \backslash\{\mSr(l^\prime)\} \cup \{\mSce(l)\}
		\Big\}$ 
        \Statex \hspace{0.8cm} (If $\mSr(l^\prime)\ = \text{ref}$, skip the loop for this particular $l'$.)
        \label{line4}
		\State Calculate the \gls{bf} vector $\bx_\text{aux}$, $U(\mS_\text{aux}),$ and $C(\mS_\text{aux})$.
		\If{$U(\mS_\text{aux})>U(\mS)$ and $C(\mS_\text{aux}) \leq \alpha \text{ (if applicable)}$}
		\State $\mS \gets \mS_\text{aux}$
		\State $\bx \gets \bx_\text{aux}$
		\EndIf
		\EndFor
		\EndFor
	\end{algorithmic}
\end{algorithm}

\subsection{Greedy AP Partitioning Algorithm} \label{sec:greedy_ap_selection}
In this subsection, a simpler iterative AP selection algorithm is introduced for all the problems. Initially, $\apref$ is selected as a reader, and a randomly selected AP is assigned as a CE. In each iteration, one AP is randomly selected and assigned to the set that yields a higher objective, i.e., $\norm{\bH_{\text{BL}} \bx}^2$ or $t$, while the \gls{dli} constraint is satisfied ($C(\mS)\leq \alpha$). If the constraint is not satisfied, the AP remains unassigned. The process continues until all APs are evaluated. This algorithm offers a lightweight alternative for the AP selection.

\section{Complexity Analysis} \label{sec:complexity}
This section provides a computational complexity analysis of the proposed algorithms.

\textbf{Channel Estimation:} For notational simplicity, we assume that $M_l = M$ and $\tau_{p,l} = \tau_p$ for all $l \in \mL \backslash  \refer$.
The complexity of line $1$ in Algorithm \ref{alg:channel_estimation} is $\mO(J' M_\refer \taupref + M_\refer^2 \taupref + M_\refer^3)$. The complexity of line $6$ is $\mO(\zeta_\text{chn}T(J' M_\refer^2 + (L-1) J' M M_\refer))$. The complexity of line $7$ is $\mO(\zeta_\text{chn}(L-1)(M_\refer M + J' M_\refer \tau_p + M_\refer^2 \tau_p + M_\refer^3))$.
The complexity of line $10$ is $\mO(\zeta_\text{chn}(L-1)J' M \tau_p + \zeta_\text{chn} L J' M_\refer \tau_p)$. The overall complexity is the summation of the given complexities above.
For $M_\refer=1$, the overall complexity reduces to $\mO(J'\taupref + (L-1)(M + J' \tau_p))$, offering a more efficient implementation and reduced hardware cost due to fewer high-resolution \glspl{adc}. For the multiple \gls{bde} case, Algorithm \ref{alg:channel_estimation} is executed for each \gls{bde}. 

\textbf{Proposed BF Designs:}
The problems $\Pbf, \Pbf^\prime,$ and $\Palpha$, given in Section \ref{sec:all_bf_sol}, have low complexities due to the closed-form solutions. Their worst-case complexity is approximately $\mO(\max(\nc,\nr) \min(\nc,\nr)^2)$, which corresponds to the cost of computing \gls{svd}. 
For $\Palpha^\prime$ in \eqref{eq:prob_Pa0_prime_convex}, the complexity is $\mO(\nc^{3.5})$ \cite{lobo1998applications}.
For the remaining problems using \gls{sdp} the complexity is $\mO(\sqrt{a} \log(1/\epsilon) (ba^3 + b^2a^2 + b^3))$, where $a \times a$ is the size of the positive semidefinite matrix, $b$ is the number of constraints, and $\epsilon$ is the solution accuracy \cite{bomze2010interior}. For example, the complexity of solving the problem in \eqref{eq:op2_real_trace2} is approximately $\mO(\sqrt{\nc} \log(1/\epsilon) (\nr \nc^3 + \nr^2 \nc^2 + \nr^3))$. The complexity of Algorithm \ref{alg:bisection} is $\mO(\log(1/\epsilon) B)$, where $\epsilon$ is the accuracy in bisection method, and $B$ is the complexity of solving the \gls{sdp} problem in \eqref{eq:feasibility_problem}.

\textbf{AP Partitioning Algorithms:} As mentioned earlier, the complexity of \gls{dp} is $\mO(LQ)$. In Algorithm \ref{alg:coalitional_game}, the primary computational cost arises from the \gls{bf} vector calculation. Therefore, the complexity of Algorithm \ref{alg:coalitional_game} is approximately $\mO(I_\text{out} L B)$, where $I_\text{out}$ is the number of outer loop iterations until convergence, and $B$ is the complexity of the \gls{bf} design.
In practice, $I_\text{out}$ is typically small, keeping the overall complexity moderate despite the algorithm's iterative nature.
Similarly, in Algorithm \ref{alg:swap_alg}, the \gls{bf} vector computation is the dominant cost, leading to an overall complexity of approximately $\mO(\abs{\mSce} \abs{\mSr} B)$.
The complexity of the greedy AP partitioning algorithm is $\mO(L B)$.

\section{Numerical Results}\label{sec:numerical_results}
\begin{figure}[tbp]
	\centering
	\includegraphics[width = 0.75\linewidth]{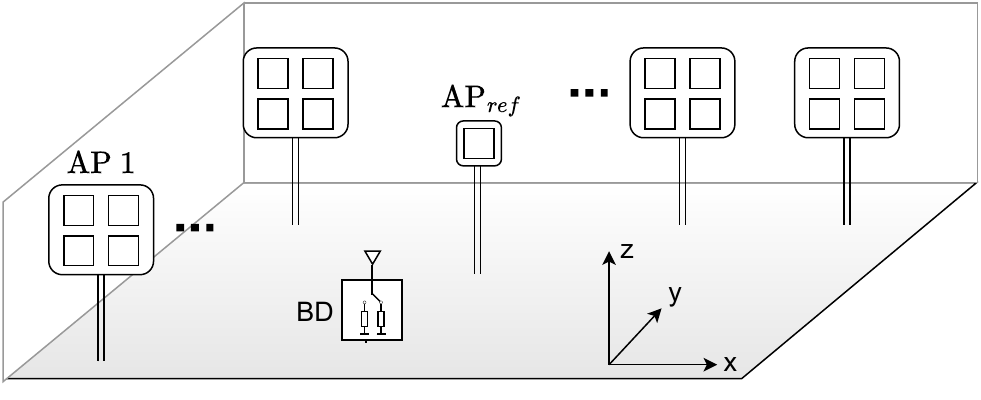}
	\caption{All \aps and \gls{bde} are located indoors with six reflectors: the side walls, the ground, and the ceiling.}
	\label{fig:simulation_environment}
\end{figure}

This section presents the simulation parameters and results.
We use the following parameters: $\lambda=0.1 \text{ m}, J=1, \gamma_j^0=-1, \gamma_j^1=1$ and the vertical and horizontal inter antenna distances are $0.5 \lambda$ for \glspl{ap}. The term $\lambda$ stands for the wavelength of the emitted signal. The solutions of the optimization problems, except for $\mP_\text{multi}$, do not depend on $P_\text{max}$. Therefore, we select $P_\text{max} = 1$ to solve these problems.
Unless otherwise stated, there are $11$ \glspl{ap} $(L=11)$, and the first $10$ \gls{ap} have $4 \times 4$ antennas in $x$$-$$z$ axis, i.e., $M_l~=~16$, $l~\in~\mL / \refer$. However, $\ap_{11}$ is the reference \gls{ap} equipped with a single antenna $(M_{11}=1)$ and $16$$-$bit \gls{adc}. 
This choice of $M_{11} = 1$ aims to achieve a system that is cost-effective and power-efficient.
While the reference \gls{ap} is located at $(10,5,2)$ \gls{m},
the remaining \glspl{ap} are distributed horizontally on the two sides of the environment at $y=1$ m and $y=9$ m. 
The center of each \gls{ap} is located at $z=2$ \gls{m}.
Unless otherwise stated, there is a single \gls{bde} located at $(4,4,2)$ \gls{m}.
The simulation environment is illustrated in Fig. \ref{fig:simulation_environment}, which considers a room of $20 \times 10 \times 4$ \gls{m} with six reflectors: four side walls, ground, and ceiling.

\begin{table}[tbp]
	\centering
	\caption{Summary of the Results for the Setup in Fig.~\ref{fig:p1}}
	\label{tab:problems2}
	\resizebox{0.49\textwidth}{!}{\begin{tabular}{|c|>{\centering\arraybackslash}m{2.1cm}|>{\centering\arraybackslash}m{1.7cm}|>{\centering\arraybackslash}m{1.7cm}|>{\columncolor{gray!14}\centering\arraybackslash}m{1.7cm}|}
			\hline
			\rowcolor[HTML]{C0C0C0} 
			\textbf{Problem} & \textbf{\gls{bf} Vector} & \textbf{PG at BD} & $\norm{\bHbl  \bx}^2$ & $C(\mS_{\text{final}})$ \\  \hline
			$\Pbf$ &  \eqref{eq:sol_mrt} & $-36.0$ dB & $-72.1$ dB & $\mathbf{58.2}$ dB \\ \hline
			$\Pdli$ & \eqref{eq:sol_dli} (\gls{cvx})  & $-34.9$ dB & $-73.5$ dB & $\mathbf{0}$ dB  \\  \hline
			$\Palpha$ & \eqref{eq:sol_bf_3} & $-35$ dB & $-73.6$ dB & $\mathbf{-261.5}$ dB \\  \hline
			$\Pbf^\prime$  & \eqref{eq:sol_for_mrt2} & $-36.2$ dB & $-52.2$ dB & $\mathbf{48.7}$ dB \\ \hline
			$\Pdli^\prime$ &  Section \ref{sec:bf_for_op2} &  $-36$ dB & $-54.8$ dB & $\mathbf{0}$ dB \\ \hline
			$\Palpha^\prime$ & Section \ref{sec:sol_for_p_a0_prime} \newline \eqref{eq:sol_Pa0_prime_closed}  & $-36.2$ dB \newline $-36.1$ dB & $-55$ dB \newline $-58.1$ dB & $\mathbf{-190.3}$ dB \newline $\mathbf{-257.1}$ dB \\ \hline
	\end{tabular}}
\end{table}

\begin{figure*}[tbp]
    \centering
    \subfloat[$\mathsf{PG}$ for $\Pbf$.]{
        \includegraphics[width=0.46\linewidth]{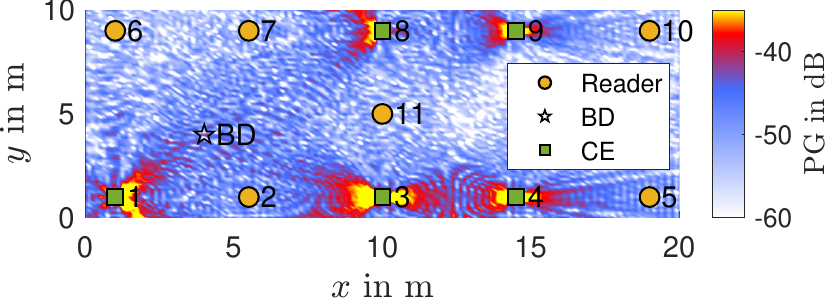}
        \label{fig:p1}
    }
    \hfil
    \subfloat[$\mathsf{PG}$ for $\Palpha$.]{
        \includegraphics[width=0.46\linewidth]{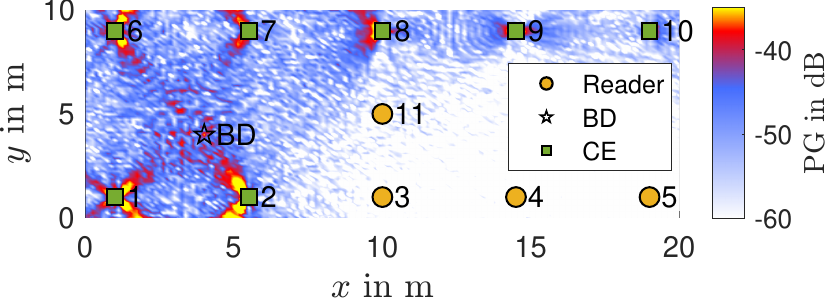}
        \label{fig:p3}
    }
    \caption{$\mathsf{PG}$ at $z=2$ m for $M_l=16, M_{11}=1$.}
    \label{fig:combined_pg}
\end{figure*}

\begin{figure*}[tbp] 
    \centering
    \subfloat[$P_\text{e}$ for $\Pbf, \Pdli,$ and $\Palpha$.]{
        \includegraphics[width=0.3\textwidth]{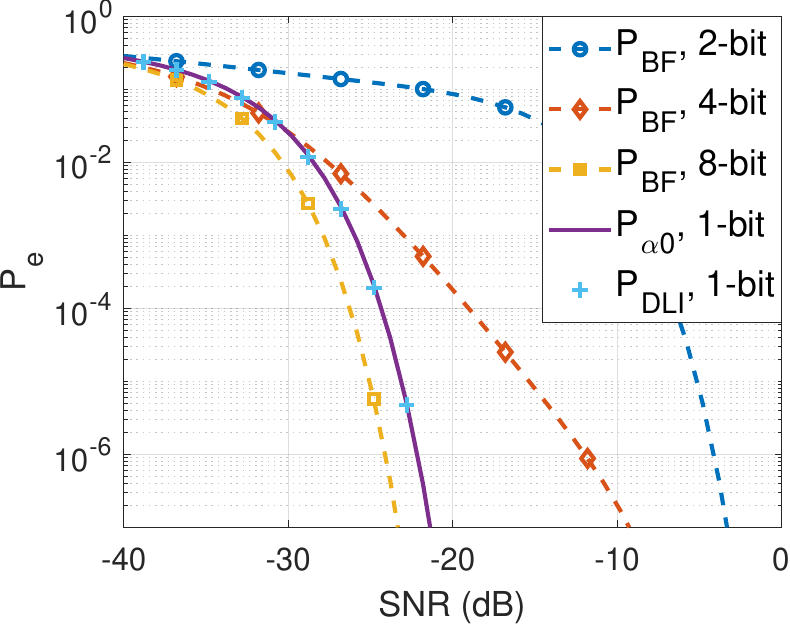}
        \label{fig:Pe_under_PCSI}
    }
    \hfil    
    \subfloat[$P_\text{e}$ for $\Pbf^\prime, \Pdli^\prime,$ and $\Palpha^\prime$.]{
        \includegraphics[width=0.3\textwidth]{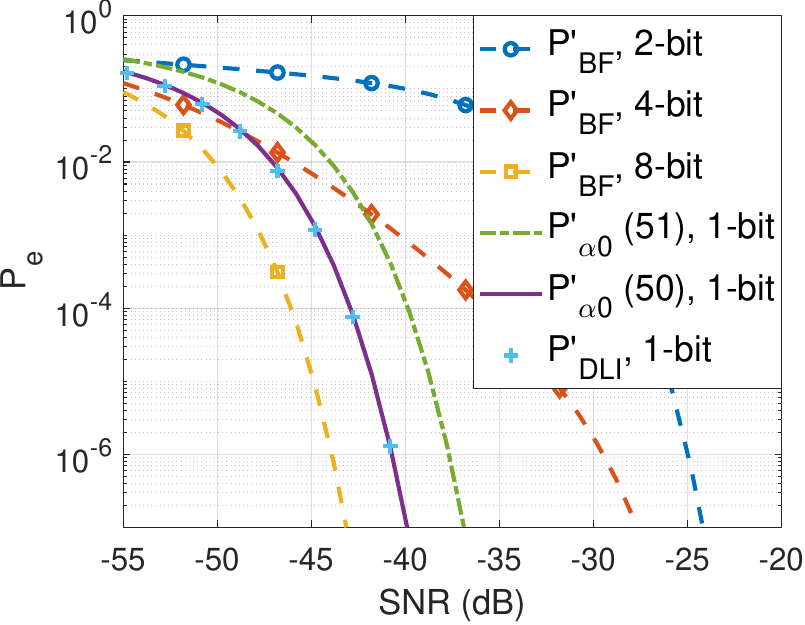}
        \label{fig:Pe_under_PCSI_prime}
    }
    \hfil
    \subfloat[Benchmark schemes for $\Palpha$.]{
        \includegraphics[width=0.3\textwidth]{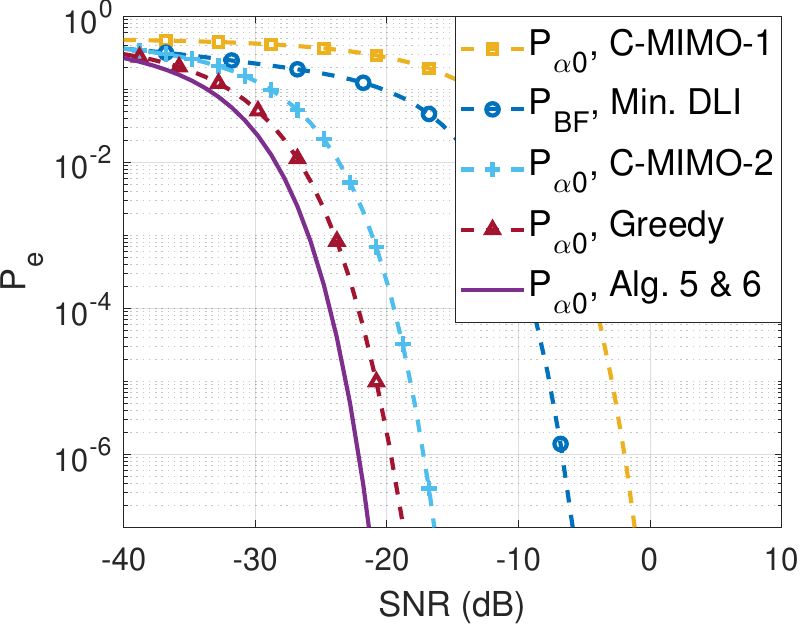}
        \label{fig:benchmarks}
    }
    \caption{Comparison of $P_\text{e}$ performance under different setups with perfect \gls{csi}.}
    \label{fig:combined_figure}
\end{figure*}

The channels are modeled as \footnote{Note that given the narrowband nature of \gls{bc} and the larger coherence bandwidth of the channel, we assume a frequency-flat channel model, despite the presence of multipath components.} 
\begin{equation}
	\allowdisplaybreaks
	\begin{split}
		[\bH_\text{DL}]_{r,c} &= \frac{\lambda}{4 \pi d_{r,c}} e^{-j \frac{2\pi}{\lambda} d_{r,c}} + \sum_{m=1}^{5} \frac{g_\text{SMC}\lambda}{4 \pi d^m_{r,c}} e^{-j \frac{2\pi}{\lambda} d_{r,c}^m}, \\
		h_{\text{C},c} &= \frac{\lambda}{4 \pi d_{c}} e^{-j \frac{2\pi}{\lambda} d_{c}} + \sum_{m=1}^{5} \frac{g_\text{SMC} \lambda}{4 \pi d^m_{c}} e^{-j \frac{2\pi}{\lambda} d_{c}^m}.
	\end{split}
\end{equation}
The distances $d_{r,c}$ and $d_c$ stand for the free-space \gls{los} path lengths between the $c$-th antenna in the \gls{ce} set and $r$-th antenna in the reader set, and $c$-th antenna in the \gls{ce} set and the \gls{bde} antenna, respectively. 
The distances $d^m_{r,c}$ and $d^m_{c}$ stands for the non-\gls{los} path lengths due to the first-order reflections.
The channel $h_{\text{R},r}$ is defined similar to $h_{\text{C},c}$, but using the distances between $r$-th antenna in the reader set and the \gls{bde} antenna, i.e., $d_{r}$ and $d^m_{r}$. 
The amplitude gain of the \glspl{smc} generated by specular reflections are $g_\text{SMC} = 0.5$.

Given that the elements of the noise vector have unit variance, the received \gls{snr} is defined as $\text{SNR}=P_\text{max} \bar{\beta}^2$, where $\bar{\beta}$ is the average path loss between an antenna of an \gls{ap} and the \gls{bde}. During the channel estimation phase, \gls{snr} is denoted as $\text{SNR}_p$.
The term $\bar{\beta} = -53.4$~dB is found by Monte-Carlo simulations for randomly distributed \gls{bde}.

Table \ref{tab:problems2} presents some of the optimization problems and the corresponding equation numbers to design the \gls{bf} vectors. While  \eqref{eq:sol_mrt}, \eqref{eq:sol_for_mrt2}, \eqref{eq:sol_bf_3},  and \eqref{eq:sol_Pa0_prime_closed} are the closed-form low-complexity solutions, the remaining solutions are obtained by solving the problems in \gls{cvx}. 
For the \gls{ap} partitioning, we use the \gls{dp} method with $s=10^{8}$ introduced in Section~\ref{sec:ap_part_for_pbf} for $\Pbf$.
For all other problems, unless stated otherwise, we use the algorithm described in Section \ref{sec:ap_partition}.
For the closed-form solutions, we set $\zeta_{\text{init}} = 30$ and $\zeta_{\text{Alg5}} = 4$. For the \gls{cvx} implementations, both parameters are set to $1$, except for $\mP_\text{multi}$, where we set $\zeta_{\text{init}} = 30$ and $\zeta_{\text{Alg5}} = 4$.

For $\Pdli$ and $\Pdli'$, 
while $\alpha=0$ dB when solving the optimization problems, we use $\alpha= 0 + \epsilon$ dB only for comparison purposes, $C(\mS) \leq 0 + \epsilon$ dB, where $\epsilon$ is a small positive number.
For $\Palpha$, $\Palpha'$, and $\mP_\text{multi}$ while $\alpha=-\infty$ dB when solving the optimization problems, we use $\alpha= -100$ dB for comparison purposes, $C(\mS) \leq -100$ dB. 
Note that the problem $\Pbf$, which uses \gls{mrt} as a \gls{bf} technique, and $\Pbf^\prime$ will serve as benchmark methods.

The path gain, $\mathsf{PG}$, represents the ratio of the received energy to the transmitted energy, calculated by
\begin{equation}
	\mathsf{PG} = \abs{\bh_\text{C}^\trp \bx}^2 /\norm{\bx}^2,
\end{equation}
where $\bh_\text{C}$ shows the channel between \glspl{ce} and the location where $\mathsf{PG}$ is calculated. 

\subsection{$\mathsf{PG}$  and $P_\text{e}$ Analyses for a Fixed Setup with Perfect \gls{csi}} \label{sec:results_for_fixed_setup}

In this subsection, all problems are analyzed for fixed \gls{ap} and \gls{bde} locations with perfect \gls{csi}.
The results are summarized in Table \ref{tab:problems2}, where $\mS_{\text{final}}$ is the final set of \glspl{ce} and readers.

Furthermore, all the results in Table \ref{tab:problems2} obtained using the proposed algorithms match the optimal solutions found through exhaustive search by checking all possible CE and reader sets combinations. While the optimality of the proposed algorithms cannot be proven, they achieve the optimal performance for the evaluated specific scenarios.

\subsubsection{Path Gain Analyses for $\mathcal{P}_{\text{\textnormal{BF}}}$ and $\Palpha$}

Figs.~\ref{fig:p1} and \ref{fig:p3} show the path gains at $z=2$ m for the problems $\Pbf$ and $\Palpha$, respectively.
In Fig.~\ref{fig:p1}, while the energy is focused on the \gls{bde} location, the readers are exposed to the \gls{dli} ($C(\mS_{\text{final}})=48.32$ dB) decreasing the performance. 
In contrast, $\Palpha$ mitigates \gls{dli} by considering it in the \gls{ap} role selection and \gls{bf} design, as shown in Fig.~\ref{fig:p3}, where $C(\mS_{\text{final}})=-259.24$ dB and energy is focused on the \gls{bde}.
This allows readers to receive the backscattered signal without interference, significantly improving performance in terms of $P_\text{e}$. 
Note that there is no need to cancel the interference in the location of $\ap_{11}$, which has high resolution \gls{adc}.

\subsubsection{$P_\text{e}$ Analysis for $\mathcal{P}_{\text{\textnormal{BF}}},\Palpha,\mathcal{P}_{\text{\textnormal{DLI}}}$}
In Fig.~\ref{fig:Pe_under_PCSI}, we compare the $P_\text{e}$ performance of $\Pbf, \Palpha,$ and $\Pdli$ with $\alpha= 1$ for the given \gls{ap} and \gls{bde} locations. We use \eqref{eq:P_e} to calculate $P_\text{e}$. 
For the $\Pbf$ case, for a higher $b$ than $b=8$, the improvement in $P_\text{e}$ will be negligible. In addition, $\Pbf$ with $b=2$ has the worst performance due to the quantization errors. 

In the \gls{cvx} solution of $\Pdli$, the final set of \glspl{ce} and readers from $\Palpha$ is used as an initial set. As shown in the figure, the performance of $\Pdli$ is nearly identical to the performance of $\Palpha$. 
This is because, as seen in Table \ref{tab:problems2}, the increase in the objective function in $\Pdli$ relative to $\Palpha$ is insufficient to produce a noticeable performance gain for the given scenario.

The proposed method $\Palpha$ with $b = 1$ outperforms $\Pbf$ with $b = 2$ and $b = 4$ due to \gls{dli} cancellation and closely matches the performance of $\Pbf$ with $b = 8$.
Therefore, $\Palpha$ is the preferred solution for the \gls{dli} cancellation due to its simplicity and optimal closed-form \gls{bf} design.

\subsubsection{$P_\text{e}$ Analysis for $\mathcal{P}'_{\text{\textnormal{BF}}},\Palpha',\mathcal{P}'_{\text{\textnormal{DLI}}}$}

In Fig.~\ref{fig:Pe_under_PCSI_prime}, we compare the $P_\text{e}$ performance of $\Pbf^\prime, \Palpha^\prime,$ and $\Pdli^\prime$ with $\alpha= 1$.
For the same error probability, there is an approximate $3$ dB \gls{snr} gap between $\Pbf^\prime, b=8$ case and $\Palpha^\prime \eqref{eq:prob_Pa0_prime_convex},b=1$ case solved using \gls{cvx}. Despite this, the slopes of the $P_\text{e}$ curves are nearly identical. Therefore, the proposed method $\Palpha^\prime$ with $b=1$ can achieve the same error probability as $\Pbf^\prime,b=8$ by increasing the transmit power, eliminating the need for power-hungry high-resolution \glspl{adc}. 

Note that the final set of \glspl{ce} and readers obtained from solving $\Palpha^\prime$ using \eqref{eq:sol_Pa0_prime_closed} are used as the initial set for the \gls{cvx} solutions of $\Palpha^\prime$ and $\Pdli^\prime$, and the performance gap between closed-form (\eqref{eq:sol_Pa0_prime_closed}) and \gls{cvx} solutions of $\Palpha^\prime$ is around $3$ dB. 
Although the \gls{cvx} solution performs better, the closed-form solution offers significantly lower complexity, making it more efficient for fast-varying channel conditions.

In addition, the performances of $\Pdli^\prime$ and $\Palpha^\prime \eqref{eq:prob_Pa0_prime_convex}$ are nearly identical because a
slight increase in the objective function of $\Pdli^\prime$ compared to $\Palpha^\prime$ does not lead to a noticeable performance improvement, as seen in Table \ref{tab:problems2}.

\begin{figure}[tbp]
	\centering
	\includegraphics[width = 0.95\linewidth]{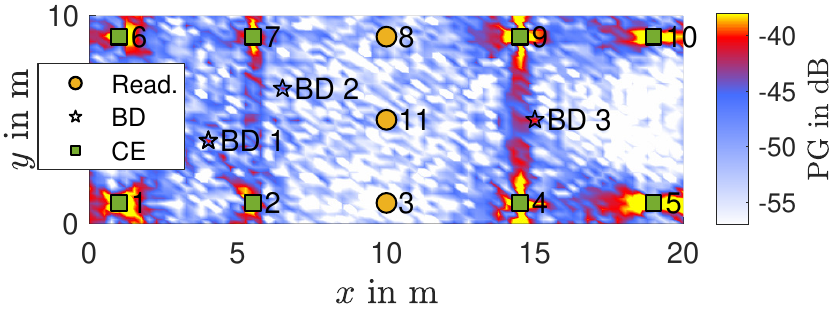}
	\caption{$\mathsf{PG}$ for $\mP_\text{multi}$.}
	\label{fig:PG_Pmulti}
\end{figure}

\subsubsection{Additional Benchmark Schemes}
Fig. \ref{fig:benchmarks} illustrates the $P_\text{e}$
calculated as in \eqref{eq:P_e} for different benchmark schemes, namely $\Palpha$ with C-MIMO-1, C-MIMO-2, and greedy, and $\Pbf$ with minimum \gls{dli}.

For the $\Pbf$ case, we set $b=2$, design $\bx$ using $(38)$, and select the AP partitioning via exhaustive search to minimize the \gls{dli}, resulting in $C\left(\mS \right)=37.8$ dB. In all other scenarios, we set $b=1$. For $\Palpha$, greedy and $\Palpha$, Alg. 5 \& 6 scenarios, we use the AP partitioning algorithm given in Section \ref{sec:greedy_ap_selection} and \ref{sec:ap_partition}, respectively.

In C-MIMO-1 and C-MIMO-2 cases, there is a single CE and a single reader, with their centers located at coordinates $(1, 5, 2)$ and $(19, 5, 2)$ m, respectively. In the C-MIMO-1 scenario, both the CE and the reader are equipped with $9 \times 9$ antennas, while in the C-MIMO-2 scenario, the CE has $12 \times 12$ antennas and the reader has $4 \times 4$ antennas. For a fair comparison, one of the reader’s antennas is equipped with a high-resolution ADC in both cases. C-MIMO-2 achieves better performance than C-MIMO-1 due to its larger number of degrees of freedom, resulting from a higher-dimensional null space of $\bHdl'$.

In addition, the proposed $\Palpha$ scheme using Algorithms 5 and 6 has the best performance compared to all the benchmarks. The $\Palpha$, greedy case provides a near-optimal solution and offers a more computationally efficient alternative.

\subsubsection{$\mathsf{PG}$ Analyses for the Multiple BDs Case} 
In this case, there are three \glspl{bde}, denoted as \gls{bde}~1, \gls{bde}~2, and \gls{bde}~3, located at $(4, 4, 2)$, $(6.5, 6.5, 2)$, and $(15, 5, 2)$ \gls{m}, respectively. The \gls{snr} is set to $10$ dB, and $\delta_k=1$. 

\begin{figure}[tbp]
	\centering
	\includegraphics[width = 0.75\linewidth]{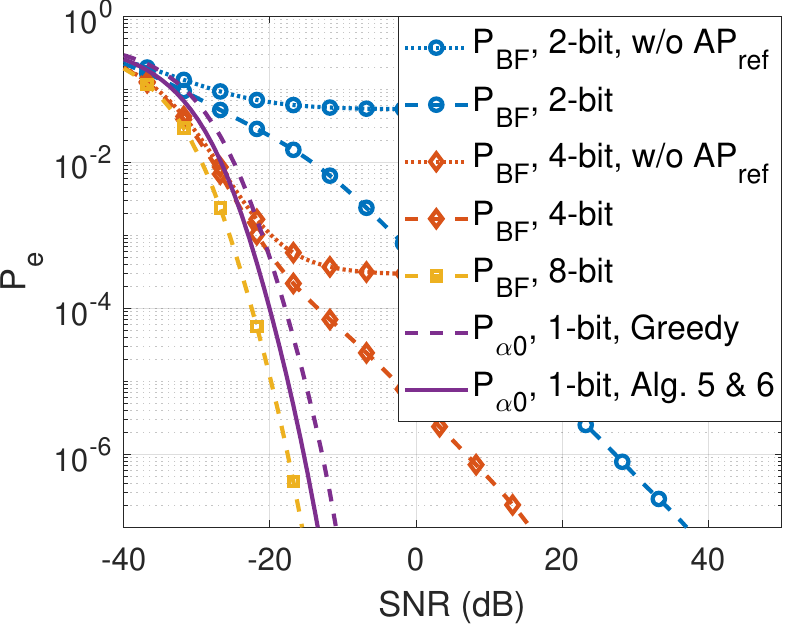}
	\caption{$P_\text{e}$ for $\Pbf$ and $\Palpha$ with/without $\apref$.}
	\label{fig:Pe_under_PCSI_montecarlo}
\end{figure}

We investigate two different scenarios. In Scenario $1$, it is assumed that \glspl{bde} use orthogonal reflection coefficients. Accordingly, we set $\bB_k=0$  and $\bC_k=0$ in \eqref{eq:SINR_cons} when solving the \gls{bf} design problem. As a result, the problem $\mP_\text{multi}$ reduces to maximizing $\min_k \norm{\bHblk \bx}^2$, subject to the transmit power and \gls{dli} constraints. Fig.~\ref{fig:PG_Pmulti} shows the resulting path gains at $z=2$ m. As shown in the figure, the \gls{dli} around the readers is mitigated ($C(\mS_{\text{final}})=-243$ dB), and energy is concentrated on the \glspl{bde}, and $\norm{\bHblk \bx}^2 / \norm{\bx}^2 = -78.3$ dB, $\forall k$.

\begin{figure*}[tbp]
    \centering
   \begin{minipage}{0.31\textwidth}
        \centering
        \includegraphics[width=\textwidth]{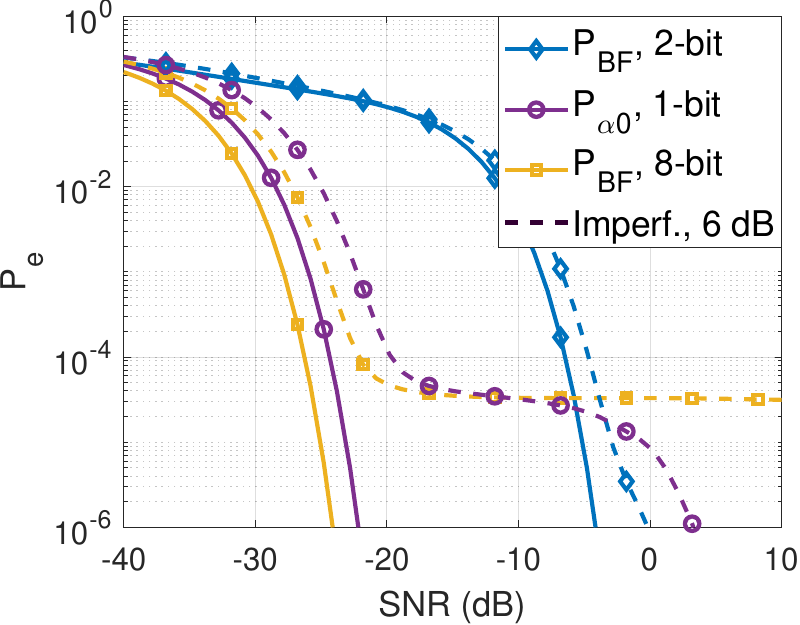}
        \caption{$P_\text{e}$ for $\Pbf$ and $\Palpha$ under imperfect \gls{csi} when $\text{SNR}_p = 6$ dB.}
        \label{fig:Pe_under_ICSI_6dB}
    \end{minipage}
    \hfill
     \begin{minipage}{0.31\textwidth}
        \centering
        \includegraphics[width=\textwidth]{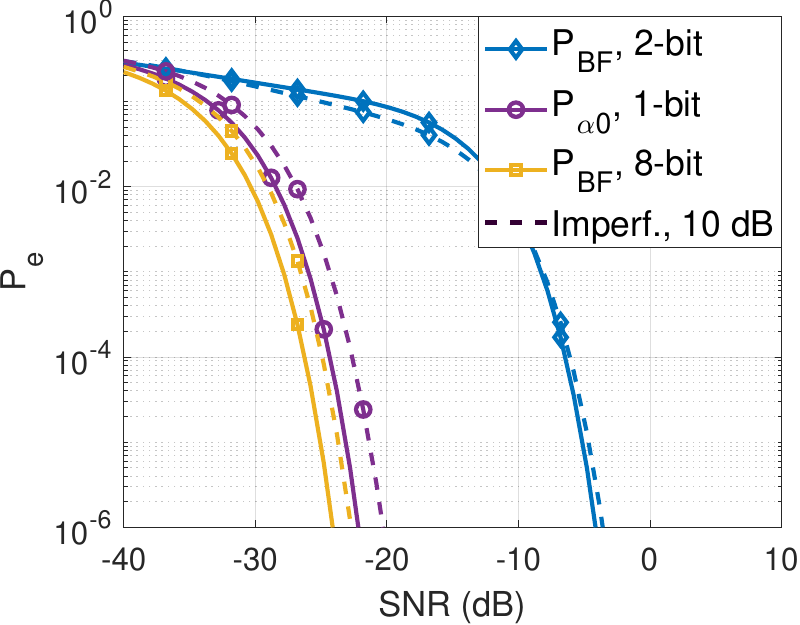}
        \caption{$P_\text{e}$ for $\Pbf$ and $\Palpha$ under imperfect \gls{csi} when $\text{SNR}_p = 10$ dB.}
        \label{fig:Pe_under_ICSI_10dB}
    \end{minipage}
    \hfill
    \begin{minipage}{0.31\textwidth}
        \centering
        \includegraphics[width=\textwidth]{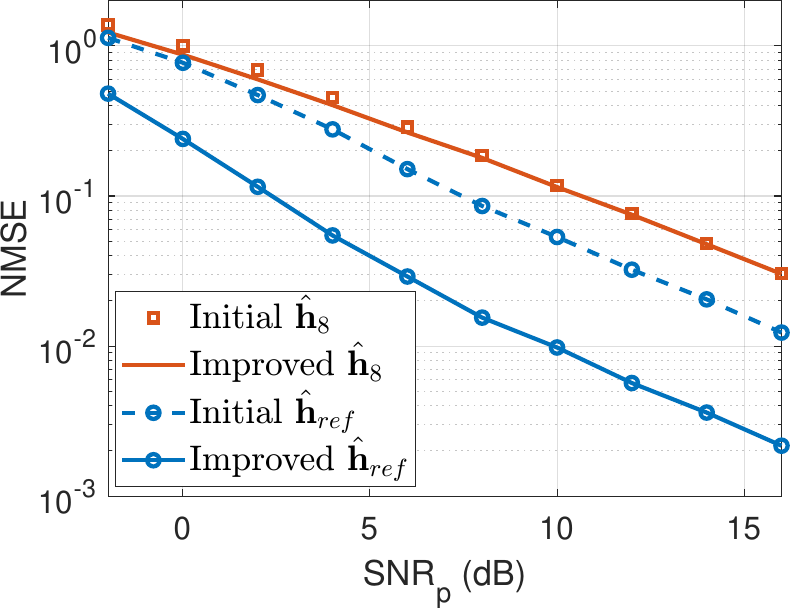}
        \caption{NMSE with and without iteration in the channel estimation.}
        \label{fig:Pe_under_ICSI_0dB}
    \end{minipage}
\end{figure*}

In Scenario 2, we compute $\bB_k$ and $\bC_k$, and consider them in \eqref{eq:SINR_cons} while maximizing the minimum \gls{sinr} across all \gls{bde}. As a result, $\text{SINR}_k = 22.7$ dB, $\norm{\bHblk \bx}^2 / \norm{\bx}^2 \approx -80 \pm 1$ dB, $\forall k$, and $C(\mS_{\text{final}})=-245$ dB.
In summary, the proposed algorithms are also effective in scenarios with multiple \glspl{bde}, achieving balanced $\text{SINR}_k$  while satisfying the \gls{dli} constraint.

\subsection{Comparison of $P_\text{e}$ for Random Setups with Perfect CSI}
In Fig. \ref{fig:Pe_under_PCSI_montecarlo}, $P_\text{e}$ is calculated using \eqref{eq:P_e} under perfect \gls{csi} for different numbers of bits in the \glspl{adc} using Monte-Carlo simulations. We analyze the cases with and without $\apref$. For without $\apref$ case, $\apref$ has the same number of \gls{adc} bits as the other \glspl{ap} and is treated as a regular \gls{ap}, eliminating the need to include it in the reader set.
\gls{csi} can be estimated without $\apref$ using geometric methods if the propagation environment, the locations of the \glspl{ap} and \gls{bde} are known \cite{deutschmann2022location}. For this reason, we present the results where $\apref$ is a regular \gls{ap}.

For each realization, there is a single \gls{bde} and its coordinate is uniformly distributed as $\mathcal{U}(0, 20)$, $\mathcal{U}(0, 10)$, and $\mathcal{U}(0, 2)$ along the $x$, $y$, and $z$ axes, respectively, while the \glspl{ap} positions remain fixed.

 In Fig. \ref{fig:Pe_under_PCSI_montecarlo}, $\Pbf$ with $b=2$ and $b=4$ w/o $\apref$ show the performance when $\apref$ is treated as a regular \gls{ap}. The performances of $\Pbf, b=8$ and $\Palpha$ without $\apref$ are not given because the performances with and without $\apref$ are almost identical.

In Fig. \ref{fig:Pe_under_PCSI_montecarlo}, the error floors in $\Pbf$ with $b=4$ and $b=2$ without $\apref$ are due to \gls{dli} and quantization noise. 
The proposed method, $\Palpha$ with $b=1$, eliminates the error floor with $1-$bit \glspl{adc}, and significantly outperforms $\Pbf$ with $b=2$ and $b=4$. Although $\Pbf$ with $b=8$ slightly outperforms $\Palpha$ with Algorithms 5 and 6, the proposed method 
achieves the same performance as $\Pbf$ with $b=8$ by increasing the \gls{snr} by approximately $2.5$ dB relative to $\Pbf$, thereby eliminating the need for high-resolution \glspl{adc}. In summary, the proposed method not only improves $P_\text{e}$ but also reduces the required \gls{adc} resolution compared to $\Pbf$.

In addition, $\Palpha$ using Algorithms 5 and 6 achieves approximately a $2$ dB SNR improvement over the greedy approach. While the greedy algorithm has a slightly lower performance, it provides a solution with reduced computational complexity.

\subsection{$P_\text{e}$ Analyses for a Fixed Setup with Imperfect \gls{csi}}
In this subsection, all problems are analyzed for fixed \glspl{ap} and the single \gls{bde} locations with imperfect \gls{csi}. While $\bHdl$ is perfectly known, all $\bh_l$s are estimated using the method in Section~\ref{sec:chn_est}. $P_\text{e}$, calculated using \eqref{eq:Pe_ICSI}, is averaged over different estimations.

Since $\apref$ has a single antenna, no iteration is required during channel estimation. The parameter $J'$ is set to $J'=2$; however, as an exception, it is increased to $J'=8$
during the estimation of $\bh_\refer$. In Figs.~\ref{fig:Pe_under_ICSI_6dB} and \ref{fig:Pe_under_ICSI_10dB}, the solid lines represent the performance under perfect \gls{csi}, while the dashed lines show the performance under imperfect \gls{csi}.

In Fig.~\ref{fig:Pe_under_ICSI_6dB}, $\text{SNR}_p$$=$$6$ dB, and poor channel estimation causes an error floor in $\Pbf, b$$=$$8$.
At high \gls{snr}, $\Pbf, b$$=$$2$ outperforms $\Pbf, b$$=$$8$ because, in the case of $\Pbf, b$$=$$2$, the diagonal element of $\hat{\bD}_j^{-1}$ corresponding to $\apref$ dominates over the other diagonal elements due to the significant \gls{adc} bit difference between $\apref$ and the remaining \glspl{ap}, as shown in \eqref{eq:variance_of_D}. Thus, the detector primarily relies on $\apref$'s signal, with minimal loss from the combination of signals from other readers. 

In Fig.~\ref{fig:Pe_under_ICSI_6dB}, a phenomenon similar to that of the $\Pbf, b$$=$$2$ case at high SNR values is also observed for $\Palpha$. While $\Palpha$ outperforms  $\Pbf$ with $b$$=$$2$ at low SNR values,
it performs worse than $\Pbf$ with $b$$=$$2$ at high \gls{snr} values, primarily due to the nullspace constraint combined with the limitations of poor channel estimation quality. In addition, $\Palpha$ offers better performance compared to the $\Pbf, b$$=$$8$ case at high \gls{snr} values. In summary, when channel estimation quality is inadequate for coherent operations of \glspl{ap}, it is more advantageous to rely solely on the received signal from a single antenna at high \gls{snr} values.
This approach eliminates the need for signal combining at the readers, thereby reducing the impact of estimation errors and ensuring more robust performance in such scenarios.

In Fig.~\ref{fig:Pe_under_ICSI_10dB}, $\text{SNR}_p$$=$$10$ dB, and the results for perfect and imperfect \gls{csi} are highly similar, indicating that the channel estimates are sufficient for the coherent operation of \glspl{ap}. Interestingly, the performance of $\Pbf, b=2$ under imperfect \gls{csi} is slightly better than that of perfect \gls{csi} in the low \gls{snr} region due to the difference in \gls{ap} role selection affecting the \gls{dli}. In summary, the proposed $\Palpha$ method offers the cost- and energy-efficient solution, leveraging the use of $1-$bit ADC.

\subsection{Performance Analyses for the Proposed Channel Estimation Algorithm}
This subsection evaluates the proposed channel estimation algorithm with fixed \glspl{ap} and \gls{bde} locations. 
$\apref$ has a $2 \times 2$ antenna array $(M_{11}=4)$ along the $x$$-$$z$ plane, and $J'$ is set to $1$. 
In Algorithm \ref{alg:gradient_descent}, the learning rate $\alpha_{lr} = 100$ and maximum iteration $T=100$, and $\zeta_{\text{chn}}=4$ in Algorithm \ref{alg:channel_estimation}.

In Fig.~\ref{fig:Pe_under_ICSI_0dB}, we compare the \gls{nmse} for $\bh_l$ with and without iterations in the channel estimation algorithm for different $\text{SNR}_p$ values. For the non-iterative case, the initial estimates of $\bh_\refer$ from Step 1 and $\bh_l$ from Step 2 are used.

The \gls{nmse} is defined as 
\begin{equation}
    \text{NMSE} = \ex{\norm{\bh_l - e^{j\theta}\hat{\bh}_l}^2 / \norm{\bh_l}^2},
\end{equation}
where the expectation is taken over $\hat{\bh}_l$. Note that all $\hat{\bh}_l$s are exposed to the same phase change due to the phase ambiguity in $\hat{\bh}_\refer$.
However, this does not impact the design of the transmitted vector $\bx$, as explained in Section~\ref{sec:estimae_of_hl}. 
Consequently, $e^{j\theta}$, where $\theta \in \{0,\pi\}$, is used to eliminate the effect of the phase ambiguity due to the estimate of $\bh_\refer$. 

In Fig.~\ref{fig:Pe_under_ICSI_0dB}, \gls{nmse} is calculated by Monte-Carlo simulations. The performance of the estimation algorithm with iteration is slightly better than that without iteration for $\hat{\bh}_8$ at low \gls{snr} values. At high \gls{snr} values, the estimates of $\bh_8$ with and without iterations become nearly identical. The results for other $\hat{\bh}_l$ values are omitted because they exhibit similar behavior to $\hat{\bh}_8$. However,  the proposed channel estimation algorithm significantly decreases \gls{nmse} for $\hat{\bh}_\refer$ across all \gls{snr} values.

\section{Conclusion} \label{sec:conclusion}
In this paper, we address the critical challenges of \gls{dli} and round-trip path loss in \gls{bibc} systems operating in distributed \gls{mimo} setups. The proposed joint \gls{ap} selection and novel \gls{bf} strategies enhance the received backscattered energy and effectively mitigate \gls{dli}, resulting in reduced error probabilities.

A tailored channel estimation algorithm is developed to tackle the impact of \gls{dli}, along with theoretical derivations for the probability of error for \gls{bc} under both perfect and imperfect \gls{csi}. The proposed iterative channel estimation algorithm improves the estimation quality compared to the initial estimates. Furthermore, the quantization noise caused by \gls{dli} is modeled to provide practical insights into system performance. 

Simulation results show that the proposed problems $(\Pdli,\Pdli',\Palpha,\Palpha', \mP_\text{multi})$ and algorithms effectively focus energy on the \gls{bde} and cancel \gls{dli}, thereby minimizing quantization noise. Using energy-efficient 1-bit \glspl{adc}, the proposed methods achieve performance comparable to benchmark setups $(\Pbf, \Pbf')$ employing higher-resolution \glspl{adc}.
Consequently, the proposed methods deliver an energy- and cost-efficient solution for \gls{bc} in multiple-antenna systems by eliminating the need for power-hungry \glspl{adc} while maintaining high performance.

\appendix

\section{Optimization Using New Objective Function} \label{apx:new_problem}

In this appendix\footnote{Note that the appendix in this arXiv version is provided as the supplementary material of the IEEE version of the paper.}, we present the formulation and solution of an alternative optimization problem, which maximizes $\norm{\bD^{-1/2} \bHbl \bx}^2$ instead of $\norm{\bHbl \bx}^2$.

We can maximize $\norm{\bD^{-1/2} \bHbl \bx}^2$ under a \emph{total} power constraint, resulting in the problem\footnote{For simplicity, the index $j$ in $\bD$ is omitted because all reflection coefficients have the same magnitude and thus $\bD$ is independent of $j$.} 
\begin{equation}
    \begin{aligned}
\mathcal{P}_{\text{D}}: \quad & \underset{\bx \in \complexset{\nc}{1}, \mS}{\text{maximize }} \quad
		& & \norm{\bD^{-1/2} \bHbl \bx}^2 \\
        & \multicolumn{1}{c}{\text{s.t.}}
		& &  \norm{\bx}^2 \leq P_\text{max}, \apref\in \mS_\text{R}.
    \end{aligned}
\end{equation} 

Alternatively, we can maximize $\norm{\bD^{-1/2} \bHbl \bx}^2$ under \emph{per-antenna} power constraints, resulting in  
\begin{equation}
    \begin{aligned}
\mathcal{P}^\prime_{\text{D}}: \quad & \underset{\bx \in \complexset{\nc}{1}, \mS}{\text{maximize }} \quad
		& & \norm{\bD^{-1/2} \bHbl \bx}^2 \\
        & \multicolumn{1}{c}{\text{s.t.}}
		& &  \abs{x_c}^2 \leq P_\text{max},  \apref\in \mS_\text{R}.
    \end{aligned}
\end{equation} 
for $c=1,\dotsc,\nc$.

Details and solutions of the new problems $\mathcal{P}_{\text{D}}$ and $\mathcal{P}^\prime_{\text{D}}$ are given below.

\subsection{Total Power Constraint ($\mathcal{P}_{\text{D}}$)}
For the given AP partitioning, the problem is
\begin{equation}
\mathcal{P}_{\text{D},1}: \underset{\bx \in \complexset{\nc}{1}}{\text{maximize }}
		 \norm{\bD^{-1/2} \bHbl \bx}^2, \text{ s.t. }
		 \norm{\bx}^2 \leq P_\text{max},
\end{equation} 
where $\bD$ is a function of $\bx$.

We employ the quadratic transform developed in \cite{shen2018fractional}
(for a different problem with the same structure). The problem $\mathcal{P}_{\text{D},1}$ is equivalent to 
\begin{equation}  \label{eq:TRS}
\underset{\bx \in \complexset{\nc}{1}, \bz \in \complexset{\nr}{1}}{\text{maximize }}
		 2\re{\bz^\herm \bHbl \bx} - \bz^\herm \bD \bz, \text{ s.t. }
		 \norm{\bx}^2 \leq P_\text{max}. 
\end{equation} 
We solve the problem by alternating optimization (AO), where $\bx$ and $\bz$ are optimized in turn until convergence.
For  given $\bx$, the optimal $\bz$ is \cite{shen2018fractional}
\begin{equation} \label{eq:y}
    \bz = \bD^{-1} \bHbl \bx.
\end{equation}
For  given $\bz$, dropping the constant terms in the objective function in \eqref{eq:TRS}, the problem can be reformulated as 
\begin{equation}\label{eq:given_y}
\mathcal{P}_{\text{D},\bx}: \underset{\bx \in \complexset{\nc}{1}}{\text{minimize }}
		 \bx^\herm \bA \bx - 2\re{\bz^\herm \bHbl \bx}, \text{ s.t. }
		 \norm{\bx}^2 \leq P_\text{max},
\end{equation}
where
\begin{equation}
\bA = \sum_{r=1}^{N_r} |z_r|^2
\frac{\bh_{\text{DL},r}^* \bh_{\text{DL},r}^\trp + \delta\, \bh_{\text{BL},r}^* \bh_{\text{BL},r}^\trp}{2^{2b_r} \times 3}
\succeq 0,
\end{equation}
for the given channel coefficients
and $z_r$ is the $r$-th element of $\bz$.
Problem  \eqref{eq:given_y} is convex and satisfies Slater’s condition (e.g., $\bx=0$
is strictly feasible), so the \gls{kkt} conditions are necessary and sufficient for optimality.
The \gls{kkt} conditions for \eqref{eq:given_y} yield the solution as
\begin{equation}
\bx = (\bA+\mu \eye)^{-1} \bHbl^\herm \bz,\quad \mu\ge 0,
\label{eq:w_lambda}
\end{equation}
where the Lagrange multiplier $\mu$ is chosen as follows: if the unconstrained solution
\(\bx = \bA^{-1} \bH_{\text{BL}}^\herm \bz\) satisfies $\|\bx\|^2 \le P_{\max}$, then $\mu = 0$; otherwise, $\mu$ is selected (e.g., by a bisection search) such that $\|\bx\|^2 = P_{\max}$.

\subsection{Per–Antenna Power Constraint ($\mathcal{P}^\prime_{\text{D}}$)}
The problem with the per-antenna transmit power constraint, for a given AP partitioning, is 
\begin{equation} 
\mathcal{P}'_{\text{D},1}: \underset{\bx \in \complexset{\nc}{1}}{\text{maximize }}
		 \norm{\bD^{-1/2} \bHbl \bx}^2, \text{ s.t. }
		 \abs{x_c}^2 \leq P_\text{max}, \forall c. 
\end{equation}
After the quadratic transform, similar to \eqref{eq:TRS}, the problem can be expressed as
\begin{equation}  \label{eq:TRSprime}
\underset{\bx \in \complexset{\nc}{1}, \bz \in \complexset{\nr}{1}}{\text{maximize }}
		 2\re{\bz^\herm \bHbl \bx} - \bz^\herm \bD \bz, \text{ s.t. }
		 \abs{x_c}^2 \leq P_\text{max}. 
\end{equation} 
The solution for $\bz$ given in Eq. \eqref{eq:y} still holds. For the given $\bz$, the problem is convex and given as
\begin{equation} \label{eq:given_y_prime}
\mathcal{P}_{\text{D},\bx}': \underset{\bx \in \complexset{\nc}{1}}{\text{minimize }}
		 \bx^\herm \bA \bx - 2\re{\bz^\herm \bHbl \bx}, \text{ s.t. }
		 \abs{x_c}^2 \leq P_\text{max},
\end{equation}
where the optimal $\bx$ can be found by \gls{pgd}. The details are as follows.

\textbf{PGD update:}
Let $g(\bx)=\bA^\trp \bx^*-\bHbl^\trp \bz^*$ be the gradient of the objective in \eqref{eq:given_y_prime} \gls{wrt} $\bx$ \cite{brandwood1983complex}. The objective is 
$L_{\bA}$-smooth with Lipschitz constant  $L_{\bA}=2\lambda_{\max}(\bA)$, where $\lambda_{\max}(\bA)$ is the maximum eigenvalue of $\bA$. With a
step size $\eta\in(0,\,1/L_{\bA}]$, 
the PGD updates are given by
\begin{equation}
\bx^{(t+1)}=\Pi_{\mathcal{C}}\left(\bx^{(t)}-\eta\, \left(g\big(\bx^{(t)}\big)\right)^*\right),
\end{equation}
where $\mathcal{C}=\{\bx:\ |x_c|^2\le P_\text{max}\}$ is the feasible set,
and 
$\Pi_{\mathcal{C}}(\cdot)$ denotes the projection onto 
$\mathcal{C}$. Since the constraint is separable across elements of 
$\bx$, the projection is performed element-wise as
\begin{equation}
\big[\Pi_{\mathcal{C}}(x)\big]_c
=\begin{cases}
x_c, & |x_c|\le \sqrt{P_\text{max}},\\[2pt]
\sqrt{P_\text{max}}\,\dfrac{x_c}{|x_c|}, & |x_c|>\sqrt{P_\text{max}}.
\end{cases}
\label{eq:projection}
\end{equation}
For convex and $L_{\bA}$-smooth objectives, PGD with $\eta\in(0,\,1/L_{\bA}]$ is guaranteed to converge to the global minimizer of the problem.

An algorithm outline for both $\mathcal{P}_{\text{D},1}$ and $\mathcal{P}_{\text{D},1}'$ is given below:
\begin{enumerate}
\item Initialize $\bx$ (pick any feasible point). 
\item Repeat until convergence:
\begin{enumerate}
\item $\bz\leftarrow \bD^{-1} \bHbl \bx$.
\item \textit{For $\mathcal{P}_{\text{D},\bx}$:} $\bx\leftarrow (\bA+\mu \eye)^{-1} \bHbl^\herm \bz$ with $\mu \geq 0$.\\
\textit{For $\mathcal{P}_{\text{D},\bx}'$:} $\bx\leftarrow$ PGD.
\end{enumerate}
\end{enumerate}

\subsection{Complexity Analysis}
This section provides a computational complexity analysis of $\mathcal{P}_{\text{D}}$ and $\mathcal{P}_{\text{D}}^\prime$ for the \gls{bf} design. 

In both problems, the AO framework iteratively updates $\bx$ and the auxiliary variable $\bz$. For each AO iteration, the complexity of $\mathcal{P}_{\text{D}}$ is approximately $\mO(\nr\nc^2 + \nc^3)$, while the complexity of $\mathcal{P}_{\text{D}}^\prime$ is $\mO(\nr\nc^2 + \frac{L_{\bA}}{\epsilon} \nc^2)$.
Here, $\mathcal{O}(L_{\bA}/\epsilon)$ denotes the number of PGD iterations required to achieve an $\epsilon$-accurate solution.

\subsection{Numerical Results}
We compare the probability of error performance of the original problems $\Palpha$ and $\Palpha'$ with that of the new problems $\mathcal{P}_{\text{D}}$ and $\mathcal{P}_{\text{D}}'$. We assume the fixed setup with perfect \gls{csi}. 

For \gls{ap} selection in $\mathcal{P}_{\text{D}}$ and $\mathcal{P}_{\text{D}}'$, we employ Algorithm 5 and Algorithm 6. As in $\Pbf^\prime$, there is no \gls{dli} constraint for the new problems $\mathcal{P}_{\text{D}}$ and $\mathcal{P}_{\text{D}}'$. Therefore, we randomly select \aps as \glspl{ce} and readers and use this initial set as input for Algorithm 5. In Algorithm 5, Phases 1 and 2 are repeated $\zeta_\text{Alg5}$ times with different initial solutions. Then, the best solution is selected.

\begin{figure}[tbp]
	\centering
	\includegraphics[width = 0.70\linewidth]{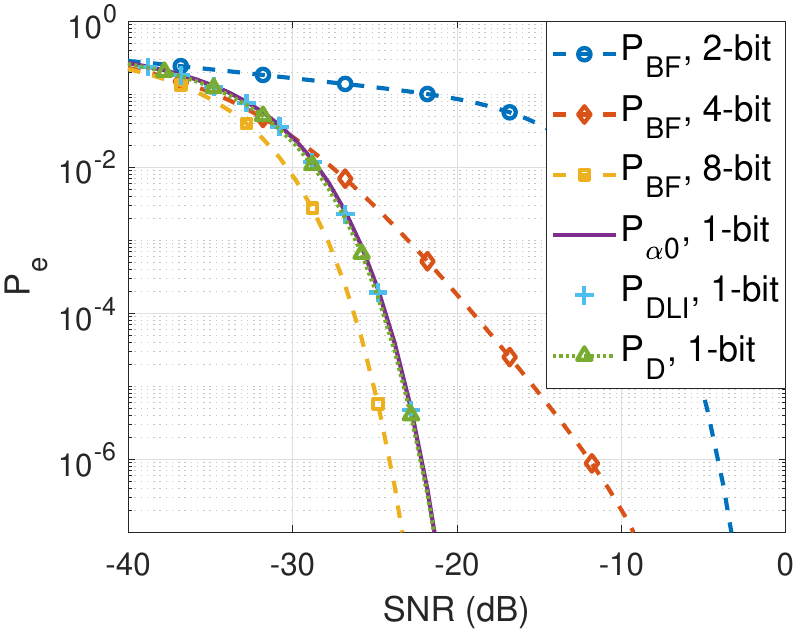}
	\caption{$P_\text{e}$ for $\Pbf, \Pdli, \Palpha$, and $\mathcal{P}_{\text{D}}$.}
	\label{fig:new_prob}
\end{figure}

\begin{figure}[tbp]
	\centering
	\includegraphics[width = 0.70\linewidth]{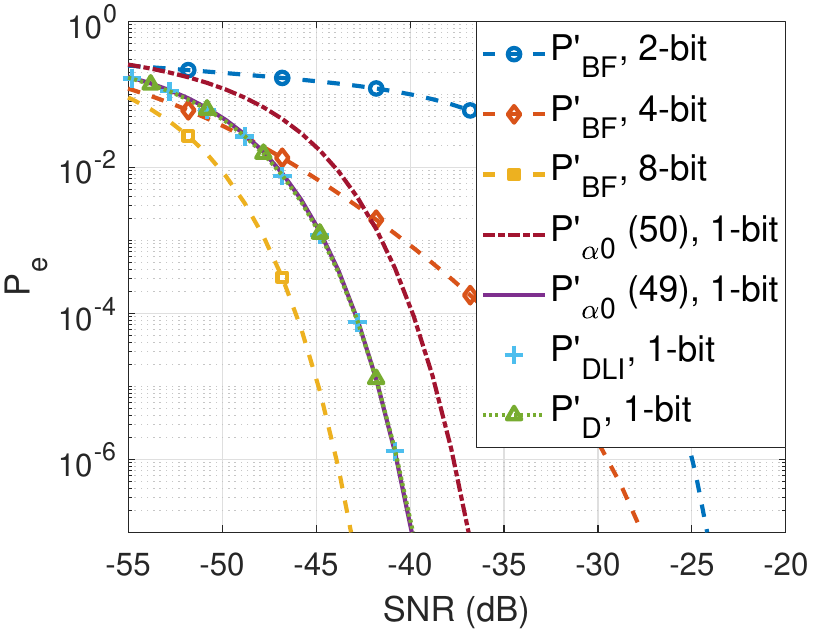}
	\caption{$P_\text{e}$ for $\Pbf^\prime, \Pdli^\prime, \Palpha^\prime$, and $\mathcal{P}^\prime_{\text{D}}$.}
	\label{fig:new_prob_prime}
\end{figure}

\begin{figure}[tbp]
	\centering
	\includegraphics[width = 0.7\linewidth]{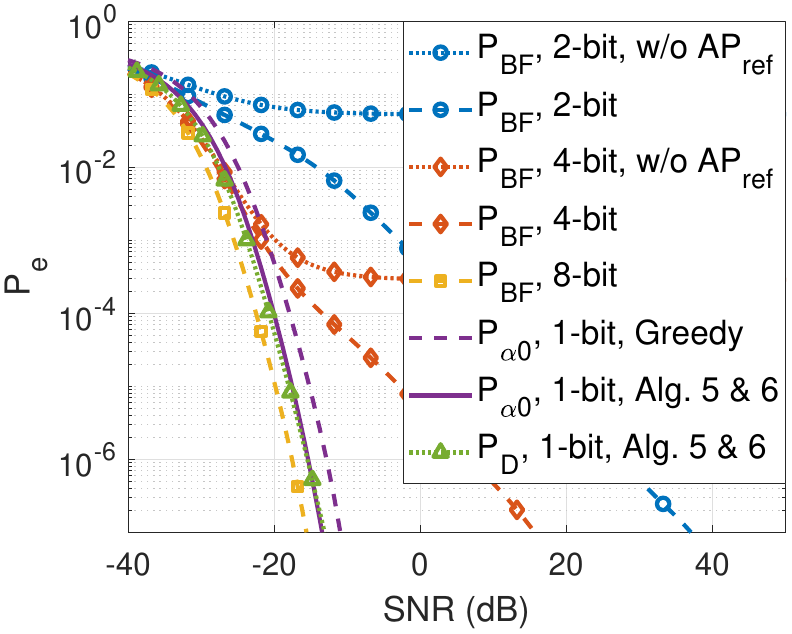}
	\caption{$P_\text{e}$ for $\Pbf, \Palpha$, and $\mathcal{P}_{\text{D}}$.}
	\label{fig:new_problem_random}
\end{figure}

Figs.~\ref{fig:new_prob} and~\ref{fig:new_prob_prime} compare the results for $\mathcal{P}_{\text{D}}$ and $\mathcal{P}_{\text{D}}'$, with the results from Figs.~\ref{fig:Pe_under_PCSI} and~\ref{fig:Pe_under_PCSI_prime}. The \aps and \gls{bde} locations are fixed. 
In Fig.~\ref{fig:new_prob}, the \gls{bf} vector for $\mathcal{P}_{\text{D}}$ is obtained by setting the \gls{snr} to $-25$ dB, while in Fig. \ref{fig:new_prob_prime}, the \gls{bf} vector for $\mathcal{P}^\prime_{\text{D}}$ is obtained by setting it to $-44.8$ dB. These \gls{bf} vectors are then used to evaluate the resulting performance over a range of \gls{snr} values. Similarly, Fig. \ref{fig:new_problem_random} is an extended version of Fig. \ref{fig:Pe_under_PCSI_montecarlo}. In this case, the optimization is performed at an \gls{snr} of $-25$ dB, and the \gls{bde} location is selected randomly.

The performance of $\Palpha$ closely matches that of $\mathcal{P}_{\text{D}}$, and likewise, the performance of $\Palpha^\prime$ is nearly identical to that of $\mathcal{P}_{\text{D}}^\prime$ in Figs.~\ref{fig:new_prob} and~\ref{fig:new_prob_prime}. In Fig.~\ref{fig:new_problem_random}, only a negligible difference is observed between the performances of $\Palpha$ and $\mathcal{P}_{\text{D}}$.
This similarity arises because, in the given setup, there are sufficient degrees of freedom to completely suppress the \gls{dli} while simultaneously focusing the power toward the \gls{bde}.

In addition, the computational complexity of $\Palpha$ is lower than that of $\mathcal{P}_{\text{D}}$. Moreover, unlike $\mathcal{P}_{\text{D}}$ and $\mathcal{P}_{\text{D}}^\prime$, which require careful selection of the \gls{snr} value during optimization, $\Palpha$ and $\Palpha'$ operate independently of any \gls{snr} information.

\bibliographystyle{IEEEtran}
\bstctlcite{IEEEexample:BSTcontrol}
\bibliography{references}
\end{document}